\documentclass[fleqn,usenatbib]{mnras}

\usepackage{newtxtext,newtxmath}

\usepackage[T1]{fontenc}

\DeclareRobustCommand{\VAN}[3]{#2}
\let\VANthebibliography\thebibliography
\def\thebibliography{\DeclareRobustCommand{\VAN}[3]{##3}\VANthebibliography}

\usepackage{graphicx}	
\usepackage{amsmath, mathtools}	

\usepackage[italicdiff]{physics}    
\usepackage{siunitx}    
\usepackage[inline, shortlabels]{enumitem} 
\usepackage[normalem]{ulem} 
\usepackage[dvipsnames]{xcolor} 

\usepackage{pgfplotstable} \pgfplotsset{compat=1.16}    
\usepackage{booktabs}   
\usepackage{caption}
\usepackage{subcaption}

\newcommand{\Msun}{\, {M_{\odot}}}

\title[GWs from a stochastically accreting neutron star]{Gravitational waves from nonradial oscillations of stochastically accreting neutron stars}

\author[W. Dong \& A. Melatos]{
Wenhao Dong,$^{1,2}$\thanks{E-mail: wddong@student.unimelb.edu.au}
Andrew Melatos,$^{1,2}$\thanks{E-mail: amelatos@unimelb.edu.au}
\\
$^{1}$School of Physics, University of Melbourne, Parkville, VIC, 3010. Australia \\
$^{2}$ARC Centre of Excellence for Gravitational Wave Discovery (OzGrav), University of Melbourne, Parkville, VIC 3010, Australia
}

\date{Accepted XXX. Received YYY; in original form ZZZ}

\pubyear{2023}

\begin{document}
\label{firstpage}
\pagerange{\pageref{firstpage}--\pageref{lastpage}}
\maketitle

\begin{abstract}
Oscillating neutron stars are sources of continuous gravitational waves. We study analytically the excitation of stellar oscillations by the mechanical impact on the stellar surface of ``clumps'' of stochastically accreted matter.
We calculate the waveform and spectrum of the gravitational wave signal emitted by the accretion-driven pulsations. 
Results are generated for an idealised model of a nonrotating, unmagnetised, one-component star with uniform polytropic index \(n_{\rm poly}\) assuming Newtonian gravity and the Cowling approximation. 
We find that the excited mode amplitudes grow with increasing \(n_{\rm poly}\) and mode order $n$. 
The gravitational wave signal forms a sequence of amplitude-modulated packets for \(n_{\rm poly}=1\), lasting \(\sim 10^{-3}\)s after each impact.
The gravitational wave strain increases with increasing \(n_{\rm poly}\), but decreases with increasing $n$ and increasing multipole order $l$ for \(n_{\rm poly}=1\).
In the observing band of current long-baseline interferometers, $g$-modes emit higher, narrower peaks in the amplitude spectral density than $f$- and $p$-modes, with the highest peaks reaching \(\sim 10^{-26}\si{Hz}^{-1/2}\) for modes with damping time \(\tau_{nl} \sim 10^{8}\si{yr}\). 
The root-mean-square strain \(h_{\text{rms}}\), calculated by summing over modes with \(2\leq l\leq4\) and \(\tau_{nl} \leq 10^{8}\si{yr}\), spans the range \(10^{-33} \leq h_{\rm rms} \leq 10^{-32}\) for \(1\leq n_{\text{poly}}\leq 2\).
\end{abstract}




\section{Introduction}

Neutron stars are likely to be sources of persistent gravitational waves (GWs), known as continuous waves \citep{Riles2023review}. General relativity predicts that continuous GWs are generated by persistent time-varying multipole moments, which may arise from nonaxisymmetric deformations such as static mountains on fast-spinning neutron stars \citep{Bildsten1998, Ushomirsky&Cutler&Bildsten2000, MelatosPayne2005, Osborne&Jones2020}, or continuously driven nonradial stellar oscillations \citep{Reisenegger&Goldreich1994, Bildsten&Cutler1995, Andersson1998a, OwenEtAl1998, Heyl2004, Lai&Wu2006, Passamonti_etal2006}.
Searches for GW signals from stellar oscillation modes analysing data from the Laser Interferometer Gravitational-wave Observatory-Virgo-KAGRA (LVK) Collaboration have focused on $r$-modes to date, assuming a quasimonochromatic signal model \citep{FesikPapa2020, MiddletonEtAl2020a, LIGO2021_SNR, LIGO2021_J0537_6910, RajbhandariEtAl2021a, CovasEtAl2022, LIGO2022_20AccretingPulsar, LIGO2022_SNR, VargasMelatos2023}. 
The searches report no detections and infer upper bounds on the characteristic wave strain at \(95\%\) confidence equivalent to \(h_{0}^{95\%} \sim 10^{-26}\), depending on the frequency band and astronomical target. 
A thorough review of historical continuous wave searches is presented by \citet{Wette2023review}.
Next-generation detectors such as the Einstein Telescope will improve the search sensitivity \citep{Maggiore_2020}.

Previous studies of neutron star oscillation modes have analysed various excitation mechanisms, including thermal excitation by type I X-ray bursts \citep{Heyl2004, Watts2012, Chambers_etal2018}, excitation during magnetar flares \citep{Levin2006, LevinVanHoven2011}, resonant tidal excitation of $g$-modes during binary inspirals \citep{Lai1994, Reisenegger&Goldreich1994, Lai&Wu2006, Sullivan_etal2023}, opacity excitation \citep{DziembowskiCassisi1999}, conversion from radial to nonradial oscillations \citep{VanHoolstEtAl1998, DziembowskiCassisi1999}, and excitation of transient superfluid modes by a rotational glitch \citep{AnderssonComer2001a, SideryEtAl2010, YimJones2020}.
Less attention has been paid to excitation by the mechanical impact of infalling matter, including stochastically accreted matter. 
Numerical simulations of radially accreting quadrupolar mass shells indicate that $f$-modes dominate the generation of gravitational radiation \citep{Nagar_etal2004}. Relatedly, \citet{deAraujo_etal2006} argued that $f$-modes in low-mass X-ray binaries are continuously excited by thermonuclear burning of accreted matter in addition to mechanical impact.
Otherwise, to the authors' knowledge, detailed calculations of neutron star oscillation modes excited by mechanical impact have not been attempted in the literature. 
Irrespective of the excitation mechanism, accretion-driven oscillations contribute to the phenomenon of spin stalling, when the gravitational radiation reaction torque balances the accretion torque \citep{Wagoner1984, Bildsten1998, WagonerEtAl2001, Riles2023review}.

In this paper, we analyse the mechanical excitation of neutron star oscillations by stochastic accretion and calculate the GW signal thereby generated. The first goal is to predict the characteristic gravitational wave strain $h_0$ as a function of the statistics of the accretion process, in order to estimate the signal's detectability. The second goal is to predict the waveform in the time and frequency domains, as a guide to the development of search algorithms for this signal class.
The paper is structured as follows. 
In Section~\ref{sec: 2-Neutron_star_model}, we describe the neutron star model and its equilibrium state.
In Section~\ref{sec: 3-Linear_adiabatic_pulsation_theory}, we briefly review the linear theory of free stellar oscillations. 
In Section~\ref{sec: 4-Accretion_excitation}, we calculate the forced oscillatory response to the mechanical impact of accreting matter using a Green's function framework established in Appendix~\ref{app:green_function_method}. We present Molleweide surface maps for the modes that are predicted to dominate the generation of gravitational radiation. 
We then compute the waveform, temporal autocorrelation function, characteristic wave strain, and amplitude spectral density of the GW signal in Sections~\ref{sec: 5-Gravitational_waveform} and \ref{sec: 6-Detectability}.
In doing so, we calculate the ensemble statistics of the GW signal as a function of the statistics of the accretion flow, e.g. the size and impact frequency of accreting ``clumps''.
Astrophysical implications are canvassed briefly in Section~\ref{sec: 7-Conclusion}.

\section{Neutron star model}
\label{sec: 2-Neutron_star_model}

The neutron star model analysed in this paper is kept simple deliberately, in order to focus attention on the novel elements of the calculation, namely mode excitation by mechanical impact and the implication for GWs. 
We assume that the star is made up of a one-component fluid obeying a polytropic equation of state (EOS). 
We also assume Newtonian gravity and zero rotation. Free, nonradial oscillations of this idealised stellar model have been analysed in detail previously by many authors \citep{Cowling1941, LedouxWalraven1958, Dziembowski1971, Cox1980}. 
We neglect the corrections introduced by a rigid crust with nonzero shear modulus \citep{McDermottEtAl1985, McDermott_etal1988, Passamonti&Andersson2012, PassamontiEtAl2021}, multiple fluid components or superfluidity \citep{McDermottEtAl1985, McDermott_etal1988, LindblomMendell1994, AnderssonComer2001, PrixRieutord2002, Passamonti&Andersson2012, GualtieriEtAl2014}, nonpolytropic EOS \citep{LindblomDetweiler1983, McDermott_etal1988, McDermott1990, Strohmayer1993}, rotation \citep{PapaloizouPringle1978, BildstenEtAl1996, Friedman&Stergioulas2013, Andersson&Gittins2023}, relativistic gravity \citep{ThorneCampolattaro1967, Thorne1969, McDermott_etal1983, Finn1988}, and magnetic fields \citep{CarrollEtAl1986, Lee2007, SotaniEtAl2007}. 
These and other corrections can be accommodated within the theoretical framework set out in Sections~\ref{sec: 3-Linear_adiabatic_pulsation_theory} and \ref{sec: 4-Accretion_excitation}, once GW detections of such sources are made. 
They will be important when analysing actual GW data but are premature for the order-of-magnitude detectability estimates in Sections~\ref{sec: 5-Gravitational_waveform} and \ref{sec: 6-Detectability}.

The free and driven oscillations analysed in Sections~\ref{sec: 3-Linear_adiabatic_pulsation_theory} and \ref{sec: 4-Accretion_excitation} respectively are linear perturbations of a nonrotating steady state which is calculated by standard methods. 
We take the EOS to be polytropic \citep{Lindblom_etal1998}, with
\begin{align}
    P(\rho) = K \rho^{1+1/n_{\text{poly}}} \,,
    \label{eqn: polytropic_EOS}
\end{align}
where \(P\) is the pressure, \(\rho\) is the density, \(K\) is a proportionality constant, and \(n_{\text{poly}}\) is the polytropic index.
The exponent \(\Gamma = \pdv*{\log{P}}{\log{\rho}} = 1+1/n_{\text{poly}}\) differs in general from the adiabatic index \(\Gamma_1\) as discussed in 
equation~\eqref{eqn: adiabatic_condition}; the latter refers to an isentropic motion without any heat exchange. The difference in \(\Gamma\) and \(\Gamma_1\) leads to $g$-modes among other phenomena as discussed in Section~\ref{subsec: 3-adiabatic_condition}.
Mass continuity implies 
\begin{align}
    \dv{M}{r} = 4\pi r^2 \rho \,,
    \label{eqn: mass_continuity}
\end{align}
where \(M(r)\) is the mass enclosed within radius $r$. 

We assume that the undisturbed star is in hydrostatic equilibrium and obeys spherical symmetry, with 
\begin{align}
    \dv{P_{0}}{r} = - \rho_{0}(r) \dv{\Phi_0}{r} \,,
    \label{eqn: hydrostatic_equilibrium}
\end{align}
where \(\Phi_0=-GM(r)/r\) is the gravitational potential, and $G$ is Newton's gravitational constant. Equilibrium quantities are denoted by the subscript zero. 
We solve \eqref{eqn: polytropic_EOS}--\eqref{eqn: hydrostatic_equilibrium} for \(\rho_{0}(r)\) subject to the standard boundary condition \(\rho_0(R_*)=0\), where \(r=R_{*}\) is the undisturbed stellar surface. 
Solutions of the Lane-Emden equations~\eqref{eqn: polytropic_EOS}--\eqref{eqn: hydrostatic_equilibrium} are well-studied \citep{Chandrasekhar1957}. 
Realistic polytropic neutron stars have \(0.5 \lesssim n_{\text{poly}} \lesssim 2\) \citep{Finn1987, LattimerPrakash2001, LattimerPrakash2016}; we study \(n_{\text{poly}}=1.0, 1.5\), and $2.0$ in this paper.
We adjust \(K\) simultaneously with \(n_{\text{poly}}\) to maintain the canonical values \(M_{*}=\SI{1.4}{\Msun}\) and \(R_{*}=\SI{10}{km}\), where \(M_{*}\) is the mass of the star.

\section{Small-amplitude stellar oscillations}
\label{sec: 3-Linear_adiabatic_pulsation_theory}

Stellar oscillations in the Newtonian regime are well-explored \citep{Dziembowski1971, Cox1980, McDermott_etal1988}. 
In this section, we summarise the main results from the theory of stellar oscillations as they apply to perturbations of the equilibrium state defined in Section~\ref{sec: 2-Neutron_star_model}. The oscillations are assumed to be linear, nonradial, and adiabatic. The equations of motion, adiabatic condition, eigenmode decomposition, and boundary conditions are presented in Sections~\ref{subsec: 3-equations_of_motion}--\ref{subsec: 3-Boundary_conditions} respectively. Examples of eigenvalues and eigenfunctions for the $f$-, $p$-, and $g$-modes relevant to the application in this paper are presented in Section~\ref{subsec: 3-numerical_results_for_eigenmodes}.

\subsection{Equations of motion}
\label{subsec: 3-equations_of_motion}

The mass and momentum conservation laws and Poisson's equation for the gravitational field are given by
\begin{align}
    \pdv{\rho}{t} + \div(\rho\vb{v}) &= 0 \,, 
    \label{eqn: continuity} \\
    \pdv{\vb{v}}{t} + (\vb{v}\vdot\grad)\vb{v} 
        &= - \frac{\grad{P}}{\rho} + \frac{1}{\rho}\div{\boldsymbol{\Sigma}} - \grad\Phi \,, 
    \label{eqn: momentum} \\
    \laplacian\Phi &= 4\pi G\rho \,, 
    \label{eqn: Poisson}
\end{align}
where \(\vb{v}, \boldsymbol{\Sigma},\) and \(\Phi\) represent the flow velocity, viscous part of the stress tensor, and gravitational potential respectively. 
Let \(\delta\) denote Eulerian perturbed quantities, e.g. \(\vb{v}=\vb{v}_0 + \delta\vb{v}\) and \(\delta\vb{v} = \pdv*{\boldsymbol{\xi}}{t} + (\vb{v}_0 \vdot \grad) \boldsymbol{\xi}\), where \(\boldsymbol{\xi}(\vb{x},t)\) is the Lagrangian displacement of a fluid element from its equilibrium position. 
Setting \(\vb{v}_0 = 0\), we write 
\begin{align}
    \delta\vb{v} = \pdv{\boldsymbol{\xi}(\vb{x},t)}{t} \,.
    \label{eqn: perturbed_velocity}
\end{align} 
The linearised versions of \eqref{eqn: continuity}--\eqref{eqn: Poisson} read 
\begin{align}
    \pdv{\delta \rho}{t} + \div(\rho_0 \delta\vb{v}) &= 0 \,,
    \label{eqn: continuity_linearised}
    \\
    \pdv{\delta\vb{v}}{t} + \frac{\grad{\delta P}}{\rho_0} 
        - \frac{\delta\rho}{\rho_{0}^2} \grad{P_0}
    &= \frac{1}{\rho_0} \div{\delta\boldsymbol{\Sigma}}
        - \frac{\delta\rho}{\rho_{0}^2} \div{\boldsymbol{\Sigma}_0} - \grad\delta\Phi 
        \,,
    \label{eqn: momentum_linearised}
    \\
    \laplacian \delta\Phi &= 4\pi G \delta\rho \,.
    \label{eqn: Poisson_linearised}
\end{align}

We simplify \eqref{eqn: continuity_linearised}--\eqref{eqn: Poisson_linearised} in two important ways in this paper. First, we neglect dissipation and set \(\boldsymbol{\Sigma} = 0\). Second, we make the Cowling approximation \citep{Cowling1941} and neglect \(\grad\delta\Phi\) with respect to the other terms in \eqref{eqn: momentum_linearised}.
With the two simplifications above, the equation of motion for \(\boldsymbol{\xi}\) becomes
\begin{align}
    0 = \pdv[2]{\boldsymbol{\xi}}{t} + \frac{\grad{\delta P}}{\rho_0} 
        - \frac{\delta\rho}{\rho_{0}^2} \grad{P_0} \,.
    \label{eqn: momentum_linearised_simplified}
\end{align}

\subsection{Adiabatic condition}
\label{subsec: 3-adiabatic_condition}

The Lagrangian perturbations for pressure and density are assumed to be related by the adiabatic condition 
\begin{align}
    \frac{\Delta P}{P_0} &= \Gamma_1 \frac{\Delta \rho}{\rho_0} \,,
    \label{eqn: adiabatic_condition}
\end{align} 
where \(\Gamma_1\) is the adiabatic index \citep{Chandrasekhar1957}. 
We adopt the convention that \(\Delta\) denotes a Lagrangian perturbation, with \(\Delta Q = \delta Q + \boldsymbol{\xi}\vdot\grad{Q}\) for a quantity $Q$. 
The adiabatic index \(\Gamma_1\) does not necessarily equal \(\Gamma\). 
We restrict attention to \(\Gamma_1 \geq \Gamma\) to avoid destabilising the $g$-modes, as discussed below.

The adiabatic, Eulerian perturbation for the pressure of an ideal fluid is 
\begin{align}
    \delta P = -\Gamma_1 P_0 \div{\boldsymbol{\xi}} - \boldsymbol{\xi}\vdot\grad{P_0} \,.
    \label{eqn: stress_tensor_perturbation}
\end{align}  
Upon substituting \eqref{eqn: continuity_linearised} and \eqref{eqn: stress_tensor_perturbation} into \eqref{eqn: momentum_linearised_simplified}, we arrive at the pulsation equation
\begin{align}
    - \pdv[2]{\boldsymbol{\xi}}{t} 
    =\, &- \grad(\Gamma_{1} \frac{P_0}{\rho_0} \div{\boldsymbol{\xi}})
        - \grad(\frac{1}{\rho_0} \boldsymbol{\xi}\vdot \grad{P_0}) 
    \nonumber \\ 
        &- \Hat{\vb{r}} \left(\frac{1}{\rho_0} \dv{\rho_0}{r} - \frac{1}{\Gamma_{1} P_0}\dv{P_0}{r}\right) 
        \Gamma_{1}\frac{P_0}{\rho_0}\div{\boldsymbol{\xi}} \,,
    \label{eqn: pulsation_equation}
\end{align}
where \(\Hat{\vb{r}}\) is the radial unit vector. 
Equation~\eqref{eqn: pulsation_equation} agrees with equation~(6) in \citet{McDermott_etal1988}, when the perturbed gravitational potential and shear modulus vanish. 

The adiabatic assumption is justified within the region \(\rho\gtrsim 10^{4}\si{\,g\,cm^{-3}}\), where the thermal conduction timescale over a pressure scale height \(P/(\rho \abs{\grad \Phi_0})\) at a temperature of \(\sim 10^8 \, {\rm K}\) is longer than the period of the surface \(l=1\) $g$-modes $(\sim 10^{-2}\,\si{s})$ \citep{Bildsten&Cutler1995}. 
For a typical polytropic EOS (\(n_{\text{poly}}=1.5\)), the non-adiabatic surface layer is thin enough (\(\sim 10^{-2}\,\si{cm}\)), that its effect on the oscillation spectrum is negligible.

\subsection{Eigenmode decomposition}
\label{subsec: 3-Eigenmode_decomposition}

The normal modes of oscillation of the linear system \eqref{eqn: pulsation_equation} exhibit a harmonic time dependence \(\boldsymbol{\xi}\propto e^{i\sigma t}\), because the coefficients of \(\boldsymbol{\xi}\) and its derivatives on the right-hand side of \eqref{eqn: pulsation_equation} do not depend on $t$.
In general, the eigenvalue \(\sigma = \omega+i\kappa\) is a complex number, whose imaginary part, \(\kappa\), describes mode damping or growth. Under adiabatic conditions (see Section~\ref{subsec: 3-adiabatic_condition}), we have \(\kappa=0\).

Two classes of oscillation modes exist, namely spheroidal and toroidal modes \citep{AizenmanSmeyers1977}. They have different angular structures. 
An inviscid, nonrotating, unmagnetised neutron star does not support toroidal motions in the Cowling approximation due to the absence of shear stress. Henceforth, we concentrate on spheroidal modes. 
These take the form
\begin{align}
    \boldsymbol{\xi} 
    = \qty[ \xi_{r}(r) Y_{lm}(\theta,\phi), \, \xi_{\perp}(r) \pdv{Y_{lm}(\theta,\phi)}{\theta}, \, \frac{\xi_{\perp}(r)}{\sin{\theta}} \pdv{Y_{lm}(\theta,\phi)}{\phi} ] e^{i\sigma t} ,
    \label{eqn: xi_decomposed}
\end{align}
where \(Y_{lm}(\theta,\phi)\) denotes a spherical harmonic \citep{Jackson1998}.

The pulsation equation~\eqref{eqn: pulsation_equation} resolves into two independent, ordinary differential equations when evaluated for the ansatz \eqref{eqn: xi_decomposed}, viz.
\begin{align}
    \dv{z_1}{\tilde{r}} 
    &= z_1 \qty(\frac{\tilde{g}}{{\tilde{c}_{s}^2}} - \frac{3}{\tilde{r}})
        + z_3 \qty[\frac{l(l+1)}{\tilde{r}} - \frac{\tilde{\sigma}^2}{\tilde{c}_{s}^2} \tilde{r}] ,
    \label{eqn: equation_of_motion_z1}
    \\
    \dv{z_3}{\tilde{r}} 
    &= z_1 \qty (\frac{1}{\tilde{r}} - \frac{\tilde{N}^2}{\tilde{\sigma}^2 \tilde{r}} ) 
        + z_3 \qty (-\tilde{A} - \frac{2}{\tilde{r}}) .
    \label{eqn: equation_of_motion_z3}
\end{align}
Equations~\eqref{eqn: equation_of_motion_z1} and \eqref{eqn: equation_of_motion_z3} are expressed in terms of the dimensionless variables
\(
z_1 = \xi_{r}/r, \,
z_3 = \xi_{\perp}/r, \,
\tilde{r} = r/R_{*}, \, 
\tilde{g} = \abs{\grad{\Phi_0}} R_{*}^2 / G M_{*}, \,
\tilde{c}_{s}^2 = c_{s}^2 R_{*} / G M_{*}, \,
\tilde{\sigma}^2 = \sigma^2 R_{*}^3 / G M_{*}
\), 
and \(
\tilde{N}^2 = N^2 R_{*}^3 / G M_{*}
\), with 
\begin{align}
    \tilde{A} &= \frac{1}{\rho_0} \dv{\rho_0}{\tilde{r}} - \frac{1}{\Gamma_{1} P_0}\dv{P_0}{\tilde{r}} \, ,
\end{align}
where \(N^2 = -\tilde{A} \abs{\grad{\Phi_0}} / R_*\) is the squared Brunt-V{\"a}is{\"a}l{\"a} frequency and \(c_s^2 = \Gamma_1 P/\rho\) is the squared adiabatic acoustic speed. $z_1$ and $z_3$ are chosen to match the notation in \citet{McDermott_etal1988}.\footnote{\citet{McDermott_etal1988} also introduced variables \(z_2\) and \(z_4\), associated with the radial and transverse tractions, which are not used in this paper.}
Other authors work with the equivalent variables \(y_1=z_1\) and \(y_2 = z_3\sigma^2 r / \abs{\grad{\Phi_0}}\) \citep{Dziembowski1971, Cox1980, McDermott_etal1988}. 

The oscillation spectrum of \eqref{eqn: equation_of_motion_z1} and \eqref{eqn: equation_of_motion_z3} consists of three modes: acoustic $p$-modes, fundamental $f$-modes, and buoyancy-driven $g$-modes. 
We follow convention and designate modes by the spherical harmonic index \(l\) and the overtone number \(n\), which counts the number of radial nodes. 
We denote $p$-modes and $g$-modes by $^lp_n$ and $^lg_n$ respectively. 
One $f$-mode exists for each \(l\), denoted by $^lf$ \citep{Cowling1941}. $f$-modes can be thought of as $p$-modes with no radial nodes. 
The azimuthal harmonic \(m\) does not appear, because the modes are degenerate with respect to \(m\), when rotation and magnetic fields are absent \citep{Perdang1968}.

\subsection{Boundary conditions}
\label{subsec: 3-Boundary_conditions}

We impose central and surface boundary conditions on equations~\eqref{eqn: equation_of_motion_z1} and \eqref{eqn: equation_of_motion_z3}. The central boundary condition follows from requiring that the Taylor expansions of \(z_1\) and \(z_3\) are finite at \(r=0\). This regularity condition reduces to 
\begin{align}
    z_1(\tilde{r}=0) - l z_3(\tilde{r}=0) = 0 \,.
    \label{eqn: inner_boundary_condition}
\end{align}
The surface boundary condition requires the Lagrangian perturbation \(\Delta P\) to vanish at the surface. This condition reduces to
\begin{align}
    z_1(\tilde{r}=1) - \tilde{\sigma}^2 z_3(\tilde{r}=1) = 0 \,.
    \label{eqn: outer_boundary_condition}
\end{align}

We normalise the amplitude of the free oscillations to the oscillatory moment of inertia \(J = M_{*}R_{*}^2\) according to
\begin{align}
    \int_V d^3x \,
    \rho \boldsymbol{\xi}^{*(\alpha)} \vdot \boldsymbol{\xi}^{(\alpha')} \, 
    &= J \delta_{\alpha \alpha'} \,,
    \label{eqn: mode_orthogonality}
\end{align}
where the volume integral is over the entire undisturbed star, \(\alpha=(n,l,m)\) labels the modes, and the superscript \({}^{*}\) denotes the complex conjugate. The orthogonality implied by equation~\eqref{eqn: mode_orthogonality} was first derived by \citet{Chandrasekhar1963}. 
The energy contained in the free oscillation labelled by \(\alpha\) under this normalisation, 
\begin{align}
    E_{\alpha} = \frac{1}{2}\sigma_{\alpha}^2 J \,,
\end{align}
scales with \(\sigma_{\alpha}^2\). 
The amplitudes of the driven modes are calculated in terms of a Green's function for excitation by accretion-driven mechanical impact in Section~\ref{sec: 4-Accretion_excitation}, while the normalisation in equation~\eqref{eqn: mode_orthogonality} is reserved for the undriven eigenfunctions. 
The energy in a mode driven by the mechanical impact of accreting gas is much smaller than \(\sigma_{\alpha}^2 J/2\) for any astrophysically plausible model of the impact process, as shown in Sections~\ref{sec: 4-Accretion_excitation}--\ref{sec: 6-Detectability}.

\subsection{Numerical results for eigenmodes}
\label{subsec: 3-numerical_results_for_eigenmodes}
\pgfplotstableread{python/data/data_with_freqHz/M1.4R1.0_G2.00n1.0L2_pmode_order0to59+freqHz.txt}\tableone
\pgfplotstableread{python/data/data_with_freqHz/M1.4R1.0_G1.67n1.5L2_pmode_order0to59+freqHz.txt}\tableoneFive
\pgfplotstableread{python/data/data_with_freqHz/M1.4R1.0_G1.67n2.0L2_pmode_order0to59+freqHz.txt}\tabletwop
\pgfplotstableread{python/data/data_with_freqHz/M1.4R1.0_G1.67n2.0L2_gmode_order1to59+freqHz.txt}\tabletwog

\pgfplotstablegetrowsof{\tableone}
\pgfmathsetmacro{\lastrowindexONE}{\pgfplotsretval}
\pgfplotstablegetrowsof{\tableoneFive}
\pgfmathsetmacro{\lastrowindexONEFive}{\pgfplotsretval}
\pgfplotstablegetrowsof{\tabletwop}
\pgfmathsetmacro{\lastrowindexTWOp}{\pgfplotsretval}
\pgfplotstablegetrowsof{\tabletwog}
\pgfmathsetmacro{\lastrowindexTWOg}{\pgfplotsretval}

\begin{table*}
    \caption{Eigenmode spectrum for three EOS with \(1 \leq n_{\text{poly}} \leq 2\) and \(5/3\leq\Gamma_1\leq2\). 
    Columns from left to right per panel: mode designations, squared eigenfrequencies $\sigma^2$ in units of $GM_*/R_*^3$, eigenfrequencies $\sigma/2\pi$ in units of Hz, eigenfunction free amplitudes (dimensionless) at the surface \(\xi_r/R_{*}\) and \(\xi_\perp/R_{*}\), dimensionless overlap integral \(Q_{nl}\), and the gravitational radiation damping time-scale \(\tau_{nl}\) in seconds for the first few $l=2$ eigenmodes. 
    We focus on $l=2$ modes as they emit quadrupole GWs, which dominate the total GW signal. 
    The free amplitudes are normalised according to the normalisation condition~\eqref{eqn: mode_orthogonality} and are much greater than the driven amplitudes resulting from an astrophysically plausible accretion process (see Sections~\ref{sec: 4-Accretion_excitation}--\ref{sec: 6-Detectability}).
    Top left: $n_{\text{poly}}=1, \Gamma_1=2$; $g$-modes are not present in this model. 
    Top right: $n_{\text{poly}}=1.5, \Gamma_1=5/3$; $g$-modes are not present in this model. 
    Bottom left and right: $n_{\text{poly}}=2, \Gamma_1=5/3$; $g$-modes are present. 
    Fixed stellar parameters: $M_{*}=1.4M_{\odot}$, $R_{*}=10^{4}$m.}
    \label{tab:M1.4_R1.0_L2}
    \begin{tabular}{@{}c@{\hspace{1cm}}c@{}}
        \multicolumn{1}{c}{$n_{\text{poly}}=1, \Gamma_1=2$} 
        & 
        \multicolumn{1}{c}{$n_{\text{poly}}=1.5, \Gamma_1=5/3$} 
        \\
        \parbox{0.45\linewidth}{
          \resizebox{\linewidth}{!}{
            \pgfplotstabletypeset[
                columns={Uorder, sigma_sq, freq_Hz, U_R, V_R, Qnl, damping}, 
                columns/Uorder/.style={string type, column name=Mode},
                columns/sigma_sq/.style={fixed zerofill, precision=3, column name=$\sigma^2 [GM_*/R_*^3]$},
                columns/freq_Hz/.style={sci, sci zerofill, sci sep align, precision=3, column name=$\sigma/2\pi$ [Hz]},
                columns/U_R/.style={fixed zerofill, precision=2, column name=$\xi_{r}(R_{*})/R_{*}$},
                columns/V_R/.style={fixed zerofill, precision=2, column name=$\xi_{\perp}(R_{*})/R_{*}$},
                columns/Qnl/.style={sci, sci zerofill, sci sep align, precision=2, column name=$Q_{n2}$},
                columns/damping/.style={sci, sci zerofill, sci sep align, precision=2, column name=$\tau_{n2}$(s)},
                col sep=space, 
                header=true,
                sci generic={mantissa sep=\times,exponent={10^{##1}}},
                every first column/.style={postproc cell content/.style={@cell content=$^2p_{##1}$}},
                every third column/.append style={
                    postproc cell content/.append code={%
                    \pgfkeysgetvalue{/pgfplots/table/@cell content}{\myTmpVal}%
                    \pgfkeysalso{my special cell/.expand once={\myTmpVal}}%
                    },
                },
                every head row/.style={before row=\midrule, after row=\midrule},
                every last row/.style={after row=\hline},
                skip rows between index={10}{\lastrowindexONE},
            ]{\tableone}
          }
        }
        &
        \parbox{0.45\linewidth}{
          \resizebox{\linewidth}{!}{
            \pgfplotstabletypeset[
                columns={Uorder, sigma_sq, freq_Hz, U_R, V_R, Qnl, damping}, 
                columns/Uorder/.style={string type, column name=Mode},
                columns/sigma_sq/.style={fixed zerofill, precision=3, column name=$\sigma^2 [GM_*/R_*^3]$},
                columns/freq_Hz/.style={sci, sci zerofill, sci sep align, precision=3, column name=$\sigma/2\pi$ [Hz]},
                columns/U_R/.style={fixed zerofill, precision=2, column name=$\xi_{r}(R_{*})/R_{*}$},
                columns/V_R/.style={fixed zerofill, precision=2, column name=$\xi_{\perp}(R_{*})/R_{*}$},
                columns/Qnl/.style={sci, sci zerofill, sci sep align, precision=2, column name=$Q_{n2}$},
                columns/damping/.style={sci, sci zerofill, sci sep align, precision=2, column name=$\tau_{n2}$(s)},
                col sep=space, 
                header=true,
                sci generic={mantissa sep=\times,exponent={10^{##1}}},
                every first column/.style={postproc cell content/.style={@cell content=$^2p_{##1}$}},
                every row 0 column 0/.style={postproc cell content/.style={@cell content=$^2f$}},
                every head row/.style={before row=\midrule, after row=\midrule},
                every last row/.style={after row=\hline},
                skip rows between index={10}{\lastrowindexONEFive},
            ]{\tableoneFive}
          }
        } \\
    \end{tabular}
    
    \vspace{0.4cm} 
    \begin{tabular}{@{}c@{\hspace{1cm}}c@{}}
        \multicolumn{1}{c}{$n_{\text{poly}}=2, \Gamma_1=5/3$} 
        & 
        \multicolumn{1}{c}{$n_{\text{poly}}=2, \Gamma_1=5/3$} 
        \\
        \parbox{0.45\linewidth}{
          \resizebox{\linewidth}{!}{
            \pgfplotstabletypeset[
                columns={Uorder, sigma_sq, freq_Hz, U_R, V_R, Qnl, damping}, 
                columns/Uorder/.style={string type, column name=Mode},
                columns/sigma_sq/.style={fixed zerofill, precision=3, column name=$\sigma^2 [GM_*/R_*^3]$},
                columns/freq_Hz/.style={sci, sci zerofill, sci sep align, precision=3, column name=$\sigma/2\pi$ [Hz]},
                columns/U_R/.style={fixed zerofill, precision=2, column name=$\xi_{r}(R_{*})/R_{*}$},
                columns/V_R/.style={fixed zerofill, precision=2, column name=$\xi_{\perp}(R_{*})/R_{*}$},
                columns/Qnl/.style={sci, sci zerofill, sci sep align, precision=2, column name=$Q_{n2}$},
                columns/damping/.style={sci, sci zerofill, sci sep align, precision=2, column name=$\tau_{n2}$(s)},
                col sep=space, 
                header=true,
                sci generic={mantissa sep=\times,exponent={10^{##1}}},
                every first column/.style={postproc cell content/.style={@cell content=$^2p_{##1}$}},
                every row 0 column 0/.style={postproc cell content/.style={@cell content=$^2f$}},
                every head row/.style={before row=\midrule, after row=\midrule},
                every last row/.style={after row=\hline},
                skip rows between index={10}{\lastrowindexTWOp},
            ]{\tabletwop}
          }
        }
        & 
        \parbox{0.45\linewidth}{
          \resizebox{\linewidth}{!}{
            \pgfplotstabletypeset[
                columns={Vorder, sigma_sq, freq_Hz, U_R, V_R, Qnl, damping}, 
                columns/Vorder/.style={string type, column name=Mode},
                columns/sigma_sq/.style={fixed zerofill, precision=4, column name=$\sigma^2 [GM_*/R_*^3]$},
                columns/freq_Hz/.style={sci, sci zerofill, sci sep align, precision=3, column name=$\sigma/2\pi$ [Hz]},
                columns/U_R/.style={fixed zerofill, precision=2, column name=$\xi_{r}(R_{*})/R_{*}$},
                columns/V_R/.style={fixed zerofill, precision=2, column name=$\xi_{\perp}(R_{*})/R_{*}$},
                columns/Qnl/.style={sci, sci zerofill, sci sep align, precision=2, column name=$Q_{n2}$},
                columns/damping/.style={sci, sci zerofill, sci sep align, precision=2, column name=$\tau_{n2}$(s)},
                col sep=space, 
                header=true,
                sci generic={mantissa sep=\times,exponent={10^{##1}}},
                every first column/.style={postproc cell content/.style={@cell content=$^2g_{##1}$}},
                every head row/.style={before row=\midrule, after row=\midrule},
                every last row/.style={after row=\hline},
                skip rows between index={10}{\lastrowindexTWOg},
            ]{\tabletwog}
          }
        } \\
    \end{tabular}
\end{table*}

We integrate equations~\eqref{eqn: equation_of_motion_z1} and \eqref{eqn: equation_of_motion_z3} using a fourth-order Runge-Kutta scheme starting from the centre and matching \eqref{eqn: outer_boundary_condition} at the surface using a Newton-Raphson shooting scheme. Eigenvalues in the short-wavelength limit (see below) serve as initial guesses. 
The approach is standard except for one subtlety:
the \(1/\tilde{r}\) terms in equations~\eqref{eqn: equation_of_motion_z1} and \eqref{eqn: equation_of_motion_z3} and the vanishing pressure and density at the surface cause \(z_1\) and \(z_3\) to diverge at \(\tilde{r}=0\) and \(\tilde{r}=1\).
The divergences are resolved by shifting the boundaries slightly. 
Specifically, we apply equations~\eqref{eqn: inner_boundary_condition} and \eqref{eqn: outer_boundary_condition} at \(\tilde{r} = \tilde{r}_1 = 10^{-10}\) and \(\rho = \rho_{B} = \SI{e7}{g\, cm^{-3}}\) respectively, noting that other physics (e.g. the magnetic stress) becomes important at \(\rho < \rho_{B}\) \citep{Bildsten&Cutler1995} and is neglected in equations~\eqref{eqn: equation_of_motion_z1} and \eqref{eqn: equation_of_motion_z3}. 
We verify numerically that the eigenvalues computed for \(\rho_{B} = \SI{e7}{g\, cm^{-3}}\) differ by one part in \(10^{6}\) from those computed for \(\rho_{B} = \SI{e4}{g\, cm^{-3}}\) (say) as a comparator, confirming that the results are insensitive to the choice of \(\rho_{B} \leq \SI{e7}{g\, cm^{-3}}\). 
Usually, about ten iterations are enough to reach the targeted accuracy \(\epsilon = \abs{\Delta\sigma}/\abs{\sigma} \leq 10^{-11}\) for $f$- and $p$-modes, and \(\epsilon = \abs{\Delta \tilde{\sigma}} \leq 10^{-11}\) for $g$-modes, where \(\Delta\sigma\) (\(\Delta\tilde{\sigma}\)) is the change in \(\sigma\) (\(\tilde{\sigma}\)) between successive iterations. 
We choose \(\epsilon\) unusually small in order to handle the cancellation arising from the oscillatory integrand in the overlap integral introduced in Section~\ref{sec: 5-Gravitational_waveform} [see equation~\eqref{eqn: Q_definition}] and Appendix~\ref{app:overlap_integral}.

\autoref{tab:M1.4_R1.0_L2} lists the eigenfrequencies calculated for $p$-, $f$-, and $g$-modes with different EOS, in order to validate the above procedure against existing results in the literature \citep[e.g.][Table 2; note the unit difference]{Robe1968}. 
$f$-modes oscillate at \(\gtrsim \si{kHz}\) depending on the EOS. 
Most $p$-modes oscillate at \(\gtrsim \SI{10}{kHz}\), except for the first overtone. 
Less centrally condensed stars (with lower $n_{\text{poly}}$) have lower frequency $f$-modes and higher-frequency overtones. The frequencies of the first ten $g$-modes run from \(\sim \SI{0.4}{kHz}\) to \(\SI{1}{kHz}\).
Note that the frequency of a $f$-mode calculated under the Cowling approximation can be 50 per cent higher than the exact value, whereas the frequencies of $g$-modes are nearly unchanged \citep{Robe1968}. Therefore, all the $g$-modes listed in the table, some $f$-modes, and the first overtone of some $p$-modes fall within the LVK observing band \citep{FinnChernoff1993, BuikemaEtAl2020}. 
The free mode amplitudes \(\xi_r\) and \(\xi_\perp\) in Table~\ref{tab:M1.4_R1.0_L2} are normalised according to \eqref{eqn: mode_orthogonality} and are much greater than the driven amplitudes resulting from an astrophysically realistic accretion process, as shown in Sections~\ref{sec: 4-Accretion_excitation}--\ref{sec: 6-Detectability}.

In the short-wavelength limit, we can obtain an analytic expression for the dispersion relation at high orders $l$ or $n$ \citep{Finn1988} through a local analysis, assuming \(\boldsymbol{\xi} \sim e^{ikr}\) and \(kR_{*}\gg 1\) \citep{McDermott_etal1983, McDermott_etal1988}. The $p$- and $g$-modes satisfy respectively 
\begin{align}
    \sigma^2_{p} &\approx k^2(r) c_{s}^2 \,, 
    \label{eqn: dispersion_relation_p} \\
    \sigma^2_{g} &\approx \frac{l(l+1)N^2}{k^2(r) r^2} \,.
    \label{eqn: dispersion_relation_g}
\end{align}
The sign of $N^2$ determines the stability of $g$-modes via \eqref{eqn: dispersion_relation_g}. For a polytropic star, one finds \(N^2\propto \Gamma_1-\Gamma < 0\) for \(\Gamma_1<\Gamma\), leading to instability. 
The values of \(\sigma_g^2 \propto  \Gamma_1-\Gamma\) in the bottom right panel of Table~\ref{tab:M1.4_R1.0_L2} are relatively low, because one has \(\Gamma_1 \approx \Gamma\) typically in a neutron star.
The slowly varying wavenumber satisfies a stationarity condition \citep{Lai1994, AertsEtAl2010}, 
\begin{align}
    \int_0^{R_{*}} dr \, k(r) = (n+C)\pi \,,
\end{align} 
where \(C\) is a constant of order unity arising from the phase change 
at the boundaries, whereupon \eqref{eqn: dispersion_relation_p} and \eqref{eqn: dispersion_relation_g} take the form
\begin{align}
    \sigma_{p, nl} &\approx (n+C)\pi \qty[ \int_0^{R_{*}} dr \, \frac{1}{c_{s}(r)} ]^{-1} \,,
    \label{eqn: asymp_freq_p} \\
    \sigma_{g, nl} &\approx \frac{\sqrt{l(l+1)}}{(n+C)\pi} \int_0^{R_{*}} dr \, \frac{N(r)}{r} \,.
    \label{eqn: asymp_freq_g}
\end{align}

Real neutron stars are viscous. Viscosity modifies the dispersion relation significantly for modes above a certain order. This happens for \(\mu\laplacian \sim \rho\pdv*{t}\), i.e. \(\mu k^2 / (4\pi^2\rho\sigma) \sim 1\) where \(\mu\) is the shear viscosity. 
For a star with temperature \(T_* = 10^{9}\)K, viscosity modifies $p$-modes with order \(n\gtrsim 10^6 (\mu/10^{25}\si{g\,cm^{-1}s^{-1}})\) and $g$-modes with order \(n\gtrsim 10^4 (\mu/10^{25}\si{g\,cm^{-1}s^{-1}})\). One gets similar estimates for the spherical harmonic order \(l\) \citep{CutlerLindblom1987, Friedman&Stergioulas2013}. We show in Section~\ref{sec: 5-Gravitational_waveform} that modes of such high order emit negligible power in GWs.

\section{Excitation by accretion}
\label{sec: 4-Accretion_excitation}

In this section, we calculate the oscillatory response of the star to the mechanical impact of gas accreting onto the stellar surface. The response is a linear combination of the eigenmodes in Section~\ref{sec: 3-Linear_adiabatic_pulsation_theory}, whose coefficients are determined by the near-surface condition at the point of impact of the accreting gas. We formulate the inhomogeneous boundary value problem in terms of a Green's function in Section~\ref{subsec: 4-Green's function}. We then solve for the response to the impact of a single clump of gas (Section~\ref{subsec: 4-Single_clump}) and a stochastic sequence of equally sized clumps with random arrival times (Section~\ref{subsec: 4-Stochastic_clumps}).

\subsection{Green's function}
\label{subsec: 4-Green's function}
We address the inhomogeneous excitation problem
\begin{align}
    \qty(\pdv[2]{t} + \mathcal{L}) \delta\vb{r} 
    = \frac{\boldsymbol{\mathcal{F}}(\vb{x},t)}{\rho_0(r)} \,,
    \label{eqn: inhomogeneous_pulsation}
\end{align}
where \(\mathcal{L}\) is the spatial operator in the pulsation equation~\eqref{eqn: pulsation_equation} (see also Appendix~\ref{app:green_function_method}) and \(\mathcal{F}_i(\vb{x},t)\) is the $i$-th Cartesian component of the mechanical force density from accretion. 
Equation~\eqref{eqn: inhomogeneous_pulsation} has the general solution 
\begin{align}
    \delta r_{k} (\vb{x},t) 
    = \int d^3x' dt'  \, G_{ki}^{*}(\vb{x},t; \vb{x'},t') \mathcal{F}_i(\vb{x}',t') \,,
    \label{eqn: General_solution_displacement}
\end{align}
where the subscript \(k\) denotes the \(k\)-th Cartesian component, the Einstein summation convention on repeated indices is adopted, and \(G_{ki}\) is the Green's function 
\begin{align}
    G_{ki}(\vb{x},t; \vb{x'},t') = H(t-t')& \sum_{\alpha} 
    \frac{\sin[\sigma_{\alpha}(t-t')]}{\sigma_{\alpha} J} 
    \xi^{*(\alpha)}_{k}(\vb{x}) \xi^{(\alpha)}_{i}(\vb{x'}) \,,
    \label{eqn: Green_function}
\end{align}
derived in Appendix~\ref{app:green_function_method}. In \eqref{eqn: Green_function}, \(H(\cdot)\) denotes the Heaviside step function, and \(\sum_\alpha\) sums over all modes.

Equation \eqref{eqn: Green_function} is equivalent to the general solution derived by \citet{SchenkEtAl2001a}, viz. equation (2.24)--(2.26) in the latter reference. The solution in \citet{SchenkEtAl2001a} follows from the mode decomposition formalism and does not introduce a Green's function explicitly. The formalism applies to both rotating and nonrotating stars. 

In this paper, we analyse the situation, where the accretion occurs stochastically through a sequence of discrete ``clumps''. Hence we model the force density from the mechanical impact of a clump at the surface phenomenologically by 
\begin{align}
    \mathcal{F}_i(\vb{x},t) = p_i\delta(\vb{x}-\vb{x}_{\text{f}}) \mathcal{T}(t) \,,
    \label{eqn: force_density general form}
\end{align}
where \(p_i\) is \(i\)-th Cartesian component of the momentum of the clump, \(\vb{x}_{\text{f}}\) is the impact point on the boundary of the star (taken to be the same for every clump for simplicity), and \(\mathcal{T}(t)\) specifies the temporal history of the impact, including its duration. 
Strictly speaking, the idealised neutron star model in this paper has a vanishing mass density at the surface; that is, there is nothing at the surface that feels the force of the impact. 
This artificial feature is resolved self-consistently by moving the outer boundary condition inward slightly, as when applying the zero-pressure boundary condition to the eigenfunctions in Section~\ref{subsec: 3-numerical_results_for_eigenmodes}, so that the star feels the force from the clump, when the clump reaches \(\rho=\rho_{\text{B}}\) (see Section~\ref{subsec: 3-numerical_results_for_eigenmodes}).

\subsection{Single clump}
\label{subsec: 4-Single_clump}
We start by considering the impact of a single clump.
If the impact occurs, such that the force density remains constant throughout the interaction between the clump and the stellar surface, we can write
\begin{align}
    \mathcal{T}(t; t_{\text{s}}) = T^{-1} \qty[H(t-t_{\text{s}}) - H(t-t_{\text{s}}-T)] \,,
    \label{eqn: tophat}
\end{align}
where \(T\) is the duration of a collision starting at \(t_{\text{s}}\). 
Upon substituting \eqref{eqn: tophat} into \eqref{eqn: General_solution_displacement} and \eqref{eqn: Green_function}, we obtain 
\begin{align}
\delta \vb{r}(t,\vb{x}; t_{\text{s}}, T) =& 
    \sum_{\alpha} 
    \frac{\vb{p} \vdot \boldsymbol{\xi}^{*(\alpha)}(\vb{x}_{\text{f}})}{\sigma_{\alpha}^2 J \, T} 
    \nonumber \\ &\times
    \left\{
        1 - \exp[-(t-t_{\text{s}}) / \tau_{\alpha}]
        \cos[\sigma_{\alpha}(t-t_{\text{s}})] 
    \right\} \,
    \boldsymbol{\xi}^{(\alpha)}(\vb{x})
    \label{eqn: displacement_tophat_lessT}
\end{align}
for \(t_{\text{s}} \leq t < t_{\text{s}}+T\), and 
\begin{align}
\delta \vb{r}(t,\vb{x}; t_{\text{s}}, T) =& 
    \sum_{\alpha} 
    \frac{\vb{p} \vdot \boldsymbol{\xi}^{*(\alpha)}(\vb{x}_{\text{f}})}{\sigma_{\alpha}^2 J \, T}
    \nonumber \\ &\times
    \{
        \exp[-(t-t_{\text{s}}-T) / \tau_{\alpha}]
        \cos[\sigma_{\alpha}(t-t_{\text{s}}-T)] 
    \nonumber \\ &
        - \exp[-(t-t_{\text{s}}) / \tau_{\alpha}] \cos[\sigma_{\alpha}(t-t_{\text{s}})] 
    \} \,
    \boldsymbol{\xi}^{(\alpha)}(\vb{x})
    \label{eqn: displacement_tophat_gtrT}
\end{align}
for \(t > t_{\text{s}}+T\).
In \eqref{eqn: displacement_tophat_lessT} and \eqref{eqn: displacement_tophat_gtrT}, we follow standard practice \citep{Echeverria1989, KokkotasEtAl2001, HoEtAl2020a} and incorporate phenomenological damping factors \(\propto \exp[-(t-t_{\text{s}})/\tau_{\alpha}]\) to capture the back reaction from gravitational radiation. A self-consistent treatment of gravitational radiation reaction lies outside the scope of this paper \citep{Wheeler1966, Thorne1969, Thorne1969b}.
We assume that the gravitational radiation reaction operates slowly over many mode cycles, viz. \(\sigma_{\alpha}\tau_{\alpha}\gg 1\), in order to simplify \eqref{eqn: displacement_tophat_lessT} and \eqref{eqn: displacement_tophat_gtrT}. 
Typical gravitational damping timescales (see Table~\ref{tab:M1.4_R1.0_L2} and Section~\ref{subsec: 5-Damping_timescale}) for $f$-modes and $^{2}p_{1}$-modes are \(\sim 10^{-2}\)s; for $g$-modes, we have \(\gtrsim10^{2}\)s. 

The factor \(\vb{p} \vdot \boldsymbol{\xi}^{(\alpha)}(\vb{x}_{\text{f}})\)
in \eqref{eqn: displacement_tophat_lessT} and \eqref{eqn: displacement_tophat_gtrT} implies that the amplitude of an excited mode depends on the angle of infall. 
$f$- and $p$-modes are more strongly excited when the clump strikes the surface radially, while $g$-modes are strongly excited, when \(\bf p\) has a sizable tangential component. For a self-consistent nonrotating picture, we take \(\vb{p} \propto \hat{\vb{r}}\), i.e. \(\vb{p}\) is always radial, unless otherwise specified.

We see from equation~\eqref{eqn: displacement_tophat_gtrT} that the oscillation resembles that excited by a delta-function impulse \(\delta(t-t_{\text{s}})\) for \(\sigma_{\alpha}T \ll 1\) and is suppressed for \(\sigma_{\alpha}T \gg 1\). 
Physically, this occurs because the top-hat impact \eqref{eqn: tophat} averages over high-frequency modes.
The height of the accretion column \(L\) is assumed to be small (\(L\lesssim\SI{1}{km}\)) for a neutron star with \(M_*=\SI{1.4}{\Msun}\) and \(R=10^4\,\si{m}\) \citep{MushtukovEtAl2018}. 
Accretion columns with \(L \sim R_*\) are possible \citep{PoutanenEtAl2013} but are not considered in this paper. 
Taking \(T\gtrsim L/c\) (see Appendix~\ref{app:duration_of_collision}), modes with frequencies \(\sigma_{\alpha}\gtrsim 10^5 \,\si{rad \,s}^{-1}\) are suppressed, i.e. \((\sigma_\alpha T)^{-1} \lesssim 1\).
However, with top-hat averaging, the system prefers to excite high-order modes, because \(\vb{p}\vdot\boldsymbol{\xi}^{*(\alpha)}(\vb{x}_{\text{f}}) \, \boldsymbol{\xi}^{(\alpha)}(\vb{x}) / (\sigma_{\alpha}^2 J\,T)\) increases with ascending order. 
This is because higher-$n$ modes require less energy to reach the same amplitude than lower-$n$ modes, as higher-$n$ modes involve larger motion in the surface layer, which is weakly bound gravitationally. 
Preferential excitation of high-order modes is also present in slowly pulsating B-type stars \citep{DupretEtAl2008, AertsEtAl2010}, as well as in Sun-like stars that are excited by turbulent convection \citep{DziembowskiEtAl2001, ChaplinEtAl2005}.\footnote{For the normalisation adopted by \citet{McDermott_etal1988} and \citet{DziembowskiEtAl2001} (\(z_1=1\) at the surface), \(J\) is replaced by \(J_\alpha\) and varies mode-by-mode.}
A more realistic collision model, e.g., the turbulent plume interaction studied by \citet{ColgatePetschek1981}, would involve extended radial and angular profiles of the force density, smooth \(\delta(\vb{x}-\vb{x}_{\text{f}})\) spatially in \eqref{eqn: force_density general form}, and average spatially the shorter-wavelength modes through \eqref{eqn: General_solution_displacement}. 
To capture this consequence qualitatively without complicating the calculation unnecessarily, we cut off the \(\alpha\) sum in \eqref{eqn: displacement_tophat_lessT} and \eqref{eqn: displacement_tophat_gtrT} at \(\abs{n} \leq n_{\text{c}} \approx 5\).
The cutoff at \(\abs{n}\approx 5\) is motivated by the computational cost required to compute the mode sum in equations~\eqref{eqn: displacement_tophat_lessT} and \eqref{eqn: displacement_tophat_gtrT} to the desired accuracy. The overlap integral [equation~\eqref{eqn: Q_expanded}] decreases monotonically with \(n\). At the same time, the grid resolution required to calculate equation~\eqref{eqn: Q_expanded} increases with \(n\). Therefore, above a certain \(n\) value, additional orders do not contribute to the desired accuracy of the mode sum. In this paper, the desired accuracy is \(0.1\%\), which implies a cutoff satisfying \(3\lesssim n \lesssim 7\), depending on the physical parameters in the problem.

\begin{figure*}
    \centering
    \includegraphics[width=\textwidth]{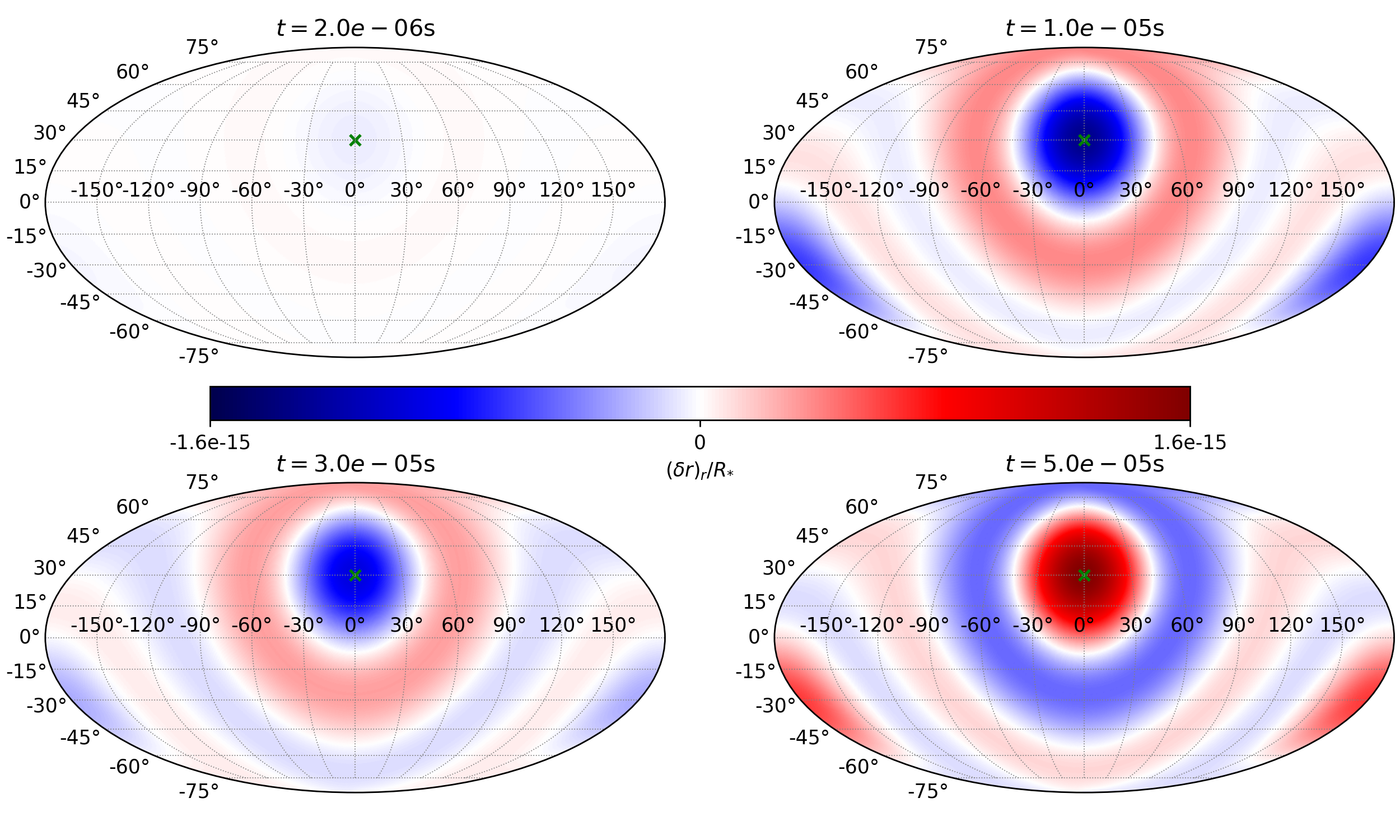}
    \caption{Molleweide plot for the (dimensionless) radial displacement of the stellar surface at different times in response to a hypothetical clump with mass $m=\SI{6e11}{kg}$ striking with speed \(\abs{\bf v} = 0.4c\) at $t=0$. The collision duration \(T\) is set to be \(T=\SI{e-5}{s}\). 
    The red and blue regions indicate outwards and inwards radial motions respectively. 
    Dark colours indicate greater displacement (absolute value), as per the colour bar.
    The impact point, marked with a cross, is at \(\theta_c = \pi/3, \phi_c = 0\). 
    $f$- and $p$-modes are summed for \(2\leq l \leq 4\). Damping is not included.
    Top left: Snapshot at $\SI{2e-6}{s}$ before the collision ends.
    Top right: Snapshot at $\SI{e-5}{s}$ exactly when the collision ends. 
    Bottom left: Snapshot at $\SI{3e-5}{s}$, three times the collision duration.
    Bottom left: Snapshot at $\SI{5e-5}{s}$, five times the collision duration.
    Stellar parameters: \(M_*=1.4\Msun, R_*=\SI{10}{km}, n_{\text{poly}}=1.5\), and \(\Gamma_1=5/3\).}
    \label{fig:NS_response_Molleweide}
\end{figure*}

Molleweide plots of the evolution of \(\delta r (r=R_*,\theta,\phi,t)\) at a sequence of four snapshots in time are presented in \autoref{fig:NS_response_Molleweide}. 
The snapshots are timed to illustrate the spreading of the oscillation. The oscillation amplitude is relatively small at the beginning as it is not fully excited before the impact ends. The surface is pushed inwards in response to the impact and bounces back within a few multiples of the collision duration. 
The striped response pattern, for \(2\leq l\leq4\), is always centred on the impact point.
The (dimensionless) radial surface displacement, 
\(\abs{\delta{r}_{\text{surf}}} / R_* = \delta{r}(t, {\bf x}, t_{\text{s}}, T) / R_*
\) for \(\abs{\bf x} = R_*\), during the spreading motion is bounded by 
\begin{align}
    \delta r_\text{surf} /R_{*} \leq
    3.9 \times 10^{-15} 
    \qty(\frac{v}{c})
    \qty(\frac{m}{\SI{6e11}{kg}})
    \qty(\frac{T}{\SI{e-5}{s}})^{-1} \,,
    \label{eqn: deltaR_scaling_single_impact}
\end{align}
where \(m\) is the mass of the clump, and \(v\) is its infall speed. 
The single-clump response is to be compared with the typical ellipticity \(\lesssim 10^{-8}\) predicted theoretically to arise from steady-state elastic and hydromagnetic mountains on the stellar surface \citep{Ushomirsky&Cutler&Bildsten2000, MelatosPayne2005, HaskellEtAl2006, PriymakEtAl2011}.

\subsection{Stochastic sequence of clumps}
\label{subsec: 4-Stochastic_clumps}
In reality, accretion onto neutron stars is a stochastic process. Complicated and possibly chaotic dynamics are imprinted on the mass accretion rate by continuously excited, three-dimensional, hydromagnetic instabilities at the disk-magnetosphere boundary \citep{RomanovaEtAl2003, RomanovaOwocki2015}. 
In this subsection, we extend the discussion about a single clump in Section~\ref{subsec: 4-Single_clump} to a random sequence of clumps. 
In the idealised phenomenological model described by \eqref{eqn: force_density general form} and \eqref{eqn: tophat}, there are several quantities which may be converted into random variables: the clump infall velocity \(\vb{v}\), the clump mass \(m\), the position \(\vb{x}_{\text{f}}\) where the accreted matter falls, and the duration of the collision \(T\).
The randomness in \(\vb{v}\) could come from turbulent or thermal motion inside the clump, three-dimensional twisting of the accretion column [e.g. magnetic Rayleigh-Taylor fingers; see \citet{RomanovaEtAl2003}], or ``bouncing'' of the stellar surface in response to previous clumps. 
However, for an accretion rate \(\dot{M} \approx 1\times10^{-8} \Msun \si{yr^{-1}}\) with mean clump impact frequency matching the LVK band, \(f_{\text{acc}} \sim {\rm kHz}\), the speed of the oscillating surface, \(v_{\text{surf}}\sim m\abs{\vb{v}} \, \abs{\boldsymbol{\xi}}^2 / (2\pi f_{\text{acc}} T \, J) \approx 10^{-15}c\), and the thermal velocity, \(v_{\text{drift}}\approx 10^{-3} (T_*/\SI{e7}{K}) c\), are negligible compared to the clump's terminal velocity, which satisfies \(\abs{\vb{v}}\sim c\).
Moreover, if the stellar magnetic field guides the clumps onto the magnetic polar cap, the impact zone is comparable to or smaller than the wavelength for most of the modes in Table~\ref{tab:M1.4_R1.0_L2} (e.g.\ with \(n\lesssim 5\)), and the factor \(\delta(\vb{x} - \vb{x}_{\text{f}})\) with fixed \(\vb{x}_{\text{f}}\) in \eqref{eqn: force_density general form} serves as a fair approximation.\footnote{
The reader is referred to \citet{KulkarniRomanova2008} and references therein for a discussion of more complicated, non-polar-cap accretion geometries.
}

The main motivation for choosing \(f_{\text{acc}} \sim 1\) kHz is to match the observing band of ground-based detectors. This value is broadly consistent with what is known about clumpy accretion streams. We expect the mean difference in the arrival times of consecutive clumps, \(f_{\text{acc}}^{-1}\), to be approximately the fragmentation period of the accretion stream. In turn, the fragmentation period is comparable to the Kepler period at the disk-magnetosphere boundary (at radius \(r_m\)), under a range of plausible physical conditions \citep{RomanovaOwocki2015}. For example, one obtains \(f_{\text{acc}}^{-1} \approx 1\)ms, for \(r_m \approx 2R_*\).
Simulations of matter fragmenting as it flows through the magnetosphere with \(10R_{*} \leq r_m \lesssim 100R_{*}\) at different accretion rates show variability on time-scales between \(\sim 10\) ms and \(\sim 100\) ms \citep[][e.g., Fig. 30 in the first reference]{RomanovaOwocki2015, MushtukovEtAl2024}.

\begin{figure*}
    \centering
    \includegraphics[width=\textwidth]{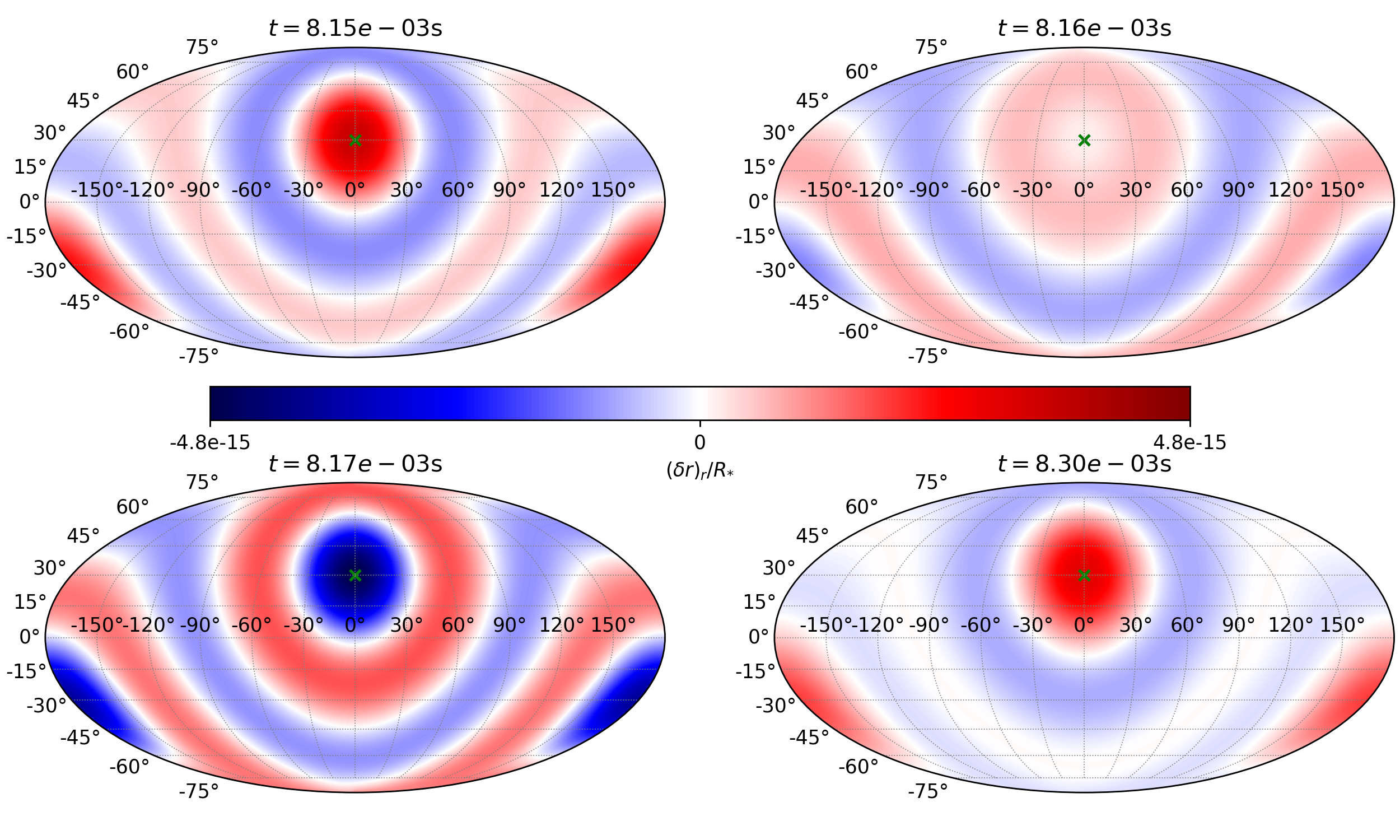}
    \caption{As for Figure~\ref{fig:NS_response_Molleweide} but for eight randomly timed clumps with the final impact lasting from \(t = 8.151\times 10^{-3}\)s to \(t = 8.161\times 10^{-3}\)s. The two top snapshots and the bottom left snapshot are deliberately chosen to depict the evolution before and after the final impact. 
    The bottom right snapshot shows a pattern of destructive interference occurring at multiple collision durations after the final impact. 
    Accretion parameters: \(\abs{\vb{v}}=0.4c, T=\SI{e-5}{s}, f_{\text{acc}}=\SI{1}{kHz}\), and \(\dot{M} = m f_{\text{acc}} = 10^{-8}\si{\Msun yr}^{-1}\).
    Stellar parameters: \(M_*=1.4\Msun, R_*=\SI{10}{km}, n_{\text{poly}}=1.5\), and \(\Gamma_1=5/3\).
    }
    \label{fig:NS_response_Molleweide_multiclumps}
\end{figure*}

In this paper, for the sake of simplicity, we assume that every clump has the same mass, and every impact has the same duration. 
We do so, because X-ray observations of accreting neutron stars are unable to resolve the impacts of individual clumps and hence measure \(m\) and \(T \sim f_{\text{acc}}^{-1}\), where the latter scaling assumes a quasi-continuous flow. Fortunately, many of the order-of-magnitude estimates in Sections~\ref{sec: 5-Gravitational_waveform} and \ref{sec: 6-Detectability} depend on \(m\) and \(f_{\text{acc}}\) chiefly through the time-averaged accretion rate \(\dot{M} = m f_{\text{acc}}\) rather than \(m\) and \(f_{\text{acc}}\) separately. It is therefore important to choose \(m\) and \(f_{\text{acc}}\) consistently with astrophysically realistic values of \(\dot{M}\).
With the above approximations, we allow the time of impact \(t_{\text{s}}\) to fluctuate randomly, i.e. 
\begin{align}
    \mathcal{F}_i = p_i\delta(\vb{x}-\vb{x}_{\text{f}}) \sum_{{\text{s}}=1}^{N(t)} \mathcal{T}(t-t_{\text{s}}) 
    \,,
    \label{eqn: shot_process}
\end{align}
where \(\{t_{\text{s}}\}\) is a sequence of impact epochs following a Poisson counting process, with mean frequency \(f_{\text{acc}}\), and \(N(t)\) is the total (and random) number of impacts up to time \(t\). 

For a canonical polytropic neutron star with \(n_{\text{poly}}=1.5\) and \(\Gamma_1 = 5/3\) accreting at \(f_{\text{acc}}\sim\si{kHz}\), the upper bound of \(\delta r_\text{surf}\) estimated using the Cauchy-Schwarz inequality is 
\((\delta r_\text{surf}/R_*)_{\text{max}} = \expval{N_{\text{clumps}}} \delta r_\text{surf}/R_*\), where \(\expval{N_{\text{clumps}}} = f_{\text{acc}} \tau\) is the mean number of clumps striking the surface during a mode-weighted average damping time \(\tau\). 
The typical value of the amplitude of \(\delta r_\text{surf}/R_*\) excited randomly by multiple clumps is expected to scale with \(\expval{N_{\text{clumps}}}^{1/2}\) and can be written as
\begin{align}
    \abs{\frac{\delta r_{\text{surf}}}{R_*}}
    \lesssim& \, 3.9\times10^{-15} 
    \expval{N_{\text{clumps}}}^{1/2}
    \qty(\frac{\dot{M}}{10^{-8} \Msun\si{yr}^{-1}}) 
    \qty(\frac{f_{\text{acc}}}{\si{kHz}})^{-1}
    \nonumber \\ &\times
    \qty(\frac{T}{10^{-5}\si{s}})^{-1} 
    \qty(\frac{\abs{\vb{v}}}{c})
    \,.
    \label{eqn: deltaR_scaling_multclumps}
\end{align}
Equations~\eqref{eqn: deltaR_scaling_multclumps} should be compared to equation~\eqref{eqn: deltaR_scaling_single_impact} for a single clump, with \(m=\dot{M}/f_{\text{acc}}\).

Figure~\ref{fig:NS_response_Molleweide_multiclumps} shows one realisation of the evolution of \(\delta r (R_*,\theta, \phi, t)\) in a sequence of four snapshots (up to \SI{8.3}{ms}) in response to eight stochastic impacts obeying \eqref{eqn: shot_process}. 
In Figure~\ref{fig:NS_response_Molleweide_multiclumps}, modes excited shortly before the second snapshot are out of phase with modes excited before the first snapshot. We see that the amplitude is lower during the impact and the oscillation spreads out. 
The third snapshot at the bottom left highlights the radially inward displacement around the impact point, shortly after the latest excitation. The last snapshot displays evidence of destructive interference of modes excited by randomly timed clumps. 
The upper bound of \(\delta r_\text{surf}/R_*\) in Figure~\ref{fig:NS_response_Molleweide_multiclumps} agrees roughly with equation~\eqref{eqn: deltaR_scaling_multclumps} for \(N_{\text{clumps}}=8\).

\section{Gravitational waveform}
\label{sec: 5-Gravitational_waveform}
In this section, we calculate the gravitational waveforms emitted by stellar oscillations excited by single and multiple clumps as in Section~\ref{sec: 4-Accretion_excitation}.
In Section~\ref{subsec: 5-Strain}, we calculate the mass quadrupole GW strain emitted by an arbitrary stellar mode (labelled by \(\alpha\), as above) in terms of an overlap integral, whose integrand is proportional to \(\boldsymbol{\xi}^{(\alpha)}\) multiplied by \(Y_{lm}\).
In Section~\ref{subsec: 5-Types_of_impact}, we calculate the waveform generated by four types of stochastic impact: single delta function, single top hat, multiple periodic top hats, and multiple Poisson top hats. Energy spectra of GW signals generated by multiple Poisson top-hat impacts with various durations are presented in Section~\ref{subsec: 5-Dominant_modes}, in order to identify the energetically dominant modes.
The GW damping timescale for each mode is calculated a posteriori in Section~\ref{subsec: 5-Damping_timescale}. 

\subsection{Strain}
\label{subsec: 5-Strain}
The metric perturbation \(h_{jk}^{\text{TT}}\) in the transverse traceless gauge measured by an observer at a distance $d$ generated by the time-varying mass multipoles of a Newtonian source is given by \citep{Thorne1980}
\begin{align}
    h_{jk}^{\text{TT}} =& \sum_{l=2}^{\infty} 
        \frac{G}{c^{2+l} d} 
            \sum_{m=-l}^{m=l} T^{\text{E}2,lm}_{jk}
            \dv[l]{I_{lm}}{t}
        \,,
    \label{eqn: strain_definition}
\end{align}
with
\begin{align}
    I_{lm} = \frac{16\pi}{(2l+1)!!} \qty[\frac{(l+1)(l+2)}{2(l-1)l}]^{1/2}
    \int d^3 x \, r^{l} Y_{lm}^{*} \delta\rho
    \,, \label{eqn: I_lm}
\end{align}
where \(t\) is the retarded time, \(\delta\rho\) is the Eulerian density perturbation from equation~\eqref{eqn: continuity_linearised}, and \(T^{\text{E2},lm}_{jk}\) is the gravitoelectric tensor spherical harmonic which describes the angular radiation pattern. 
Only the mass multipoles emit GWs in a nonrotating and inviscid neutron star, because the current multipoles \(\propto \curl[\rho_0 \boldsymbol{\xi}^{(\alpha)}]\) have no radial component for spheroidal modes. 

The time derivative in \eqref{eqn: strain_definition}, after integrating \eqref{eqn: I_lm} by parts, can be written as 
\begin{align}
    \dv[l]{I_{lm}}{t}
    =& \sum_{n} Q_{nl} M_{*} R_{*}^{l}
    \dv[l]{t} \sum_{{\text{s}}=1}^{N(t)} A_{\alpha}(t; t_{\text{s}}) \,,
\end{align}
plus the time derivative of a surface integral which vanishes for \(\rho(R_*)=0\). 
In the above notation, we have 
\begin{align}
    A_{\alpha}(t; t_{\text{s}}) = 
    \frac{\vb{p} \vdot \boldsymbol{\xi}^{*(\alpha)}(\vb{x}_\text{f})}{J}
    \int dt' \, g_{\alpha}(t-t') \mathcal{T}(t'-t_{\text{s}}) \,,
    \label{eqn: A_alpha(t;t_s)}
\end{align}
which equals the mode amplitude excited by a clump striking at time \(t_{\text{s}}\). We also have
\begin{align}
    g_{\alpha}(t) &= 
    \frac{H(t) \sin(\sigma_{\alpha}t) \exp(-t / \tau_{\alpha}) }{\sigma_{\alpha}} 
    \,,
\end{align}
while \(Q_{nl}\) is the dimensionless overlap integral \citep{PressTeukolsky1977, Lai1994, Reisenegger&Goldreich1994, KokkotasSchafer1995, Sullivan_etal2023} 
\begin{align}
    Q_{nl} &= \frac{1}{M_{*}R_{*}^{l}} \int d^3 x \, 
        \rho_0 \boldsymbol{\xi}^{*(\alpha)}(\vb{x}) \vdot \grad[r^{l}Y_{lm}(\theta,\phi)]
    \label{eqn: Q_definition}
    \\
    &= \frac{l}{M_{*}R_{*}^{l}} \int_{0}^{R_{*}} dr \,
        \rho_0(r) r^{l+1} \qty[\xi_{r}^{(nl)}(r) + (l+1) \xi_{\perp}^{(nl)}(r)] \,.
    \label{eqn: Q_expanded}
\end{align}
The values of \(Q_{n2}\) for the first few $n$ values for different EOS are quoted in Table~\ref{tab:M1.4_R1.0_L2}. 
The table confirms that $f$-modes emit GWs most efficiently, in the sense that $f$-modes have the largest \(Q_{n2}\).
Appendix~\ref{app:overlap_integral} discusses how to calculate \(Q_{nl}\) numerically, a nontrivial task.

The oscillatory character of the integrand in \eqref{eqn: Q_expanded} causes \(\abs{Q_{nl}}\) to decrease, as the number of radial nodes \(n\) increases. 
An empirical relation between \(\abs{Q_{nl}}\) and \(n\) is presented in Appendix~\ref{app:overlap_integral}. In this paper, we restrict GW calculations to modes with \(n \leq n_{\text{c}} \approx 5\) (see Section~\ref{subsec: 4-Single_clump}), for which one has \(\abs{Q_{nl}} > 10^{-3} \max_{n'} \abs{Q_{n'l}}\).

\begin{figure*}
    \centering
    \begin{subfigure}{\columnwidth}
        \centering
        \includegraphics[width=\textwidth]{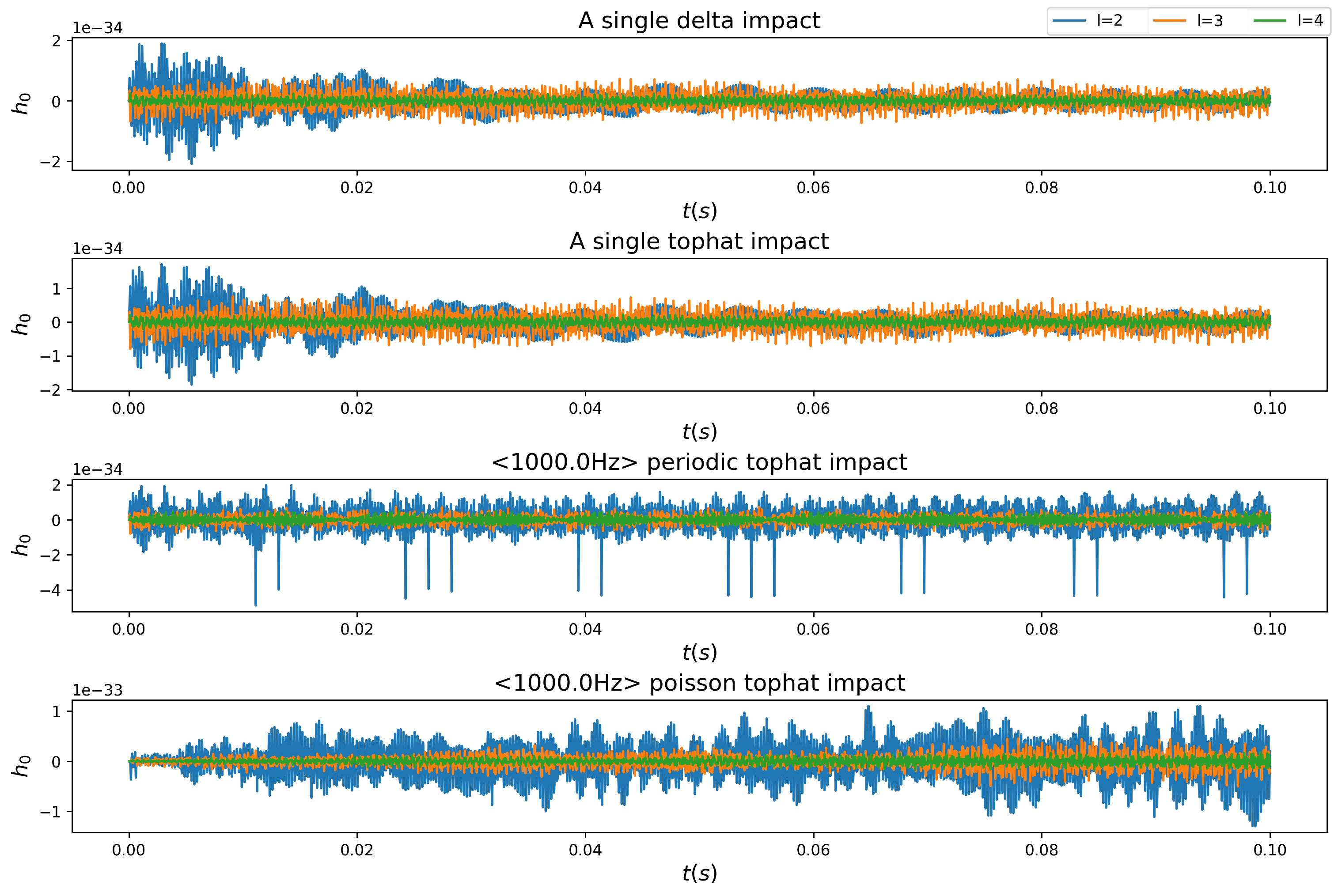}
        \caption{$n_{\text{poly}}=1, \Gamma_1=2$}
    \end{subfigure}
    \hfill
    \begin{subfigure}{\columnwidth}
        \includegraphics[width=\textwidth]{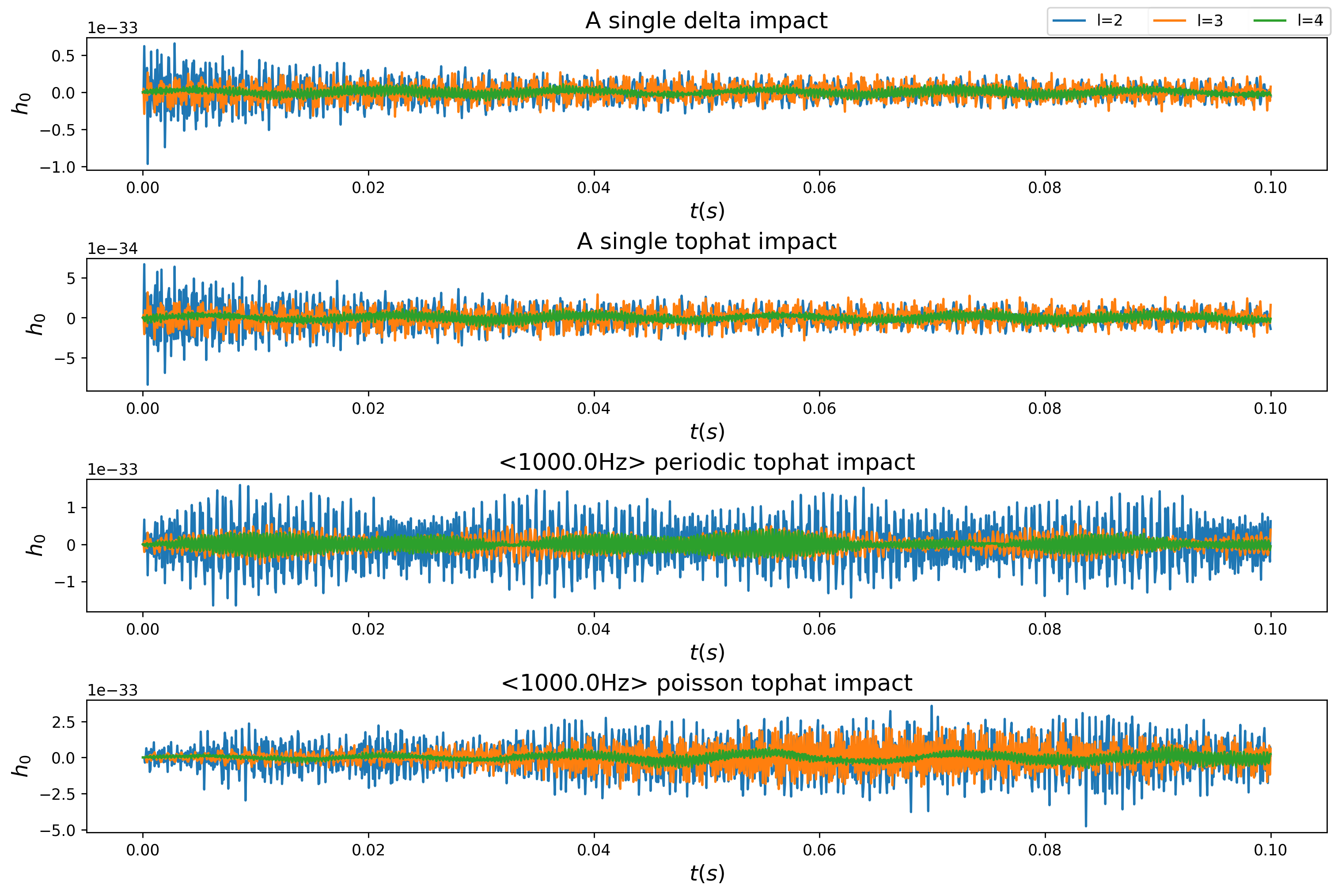}
        \caption{$n_{\text{poly}}=1.5, \Gamma_1=5/3$}
    \end{subfigure}
    \vfill
    \begin{subfigure}{\columnwidth}
        \centering
        \includegraphics[width=\textwidth]{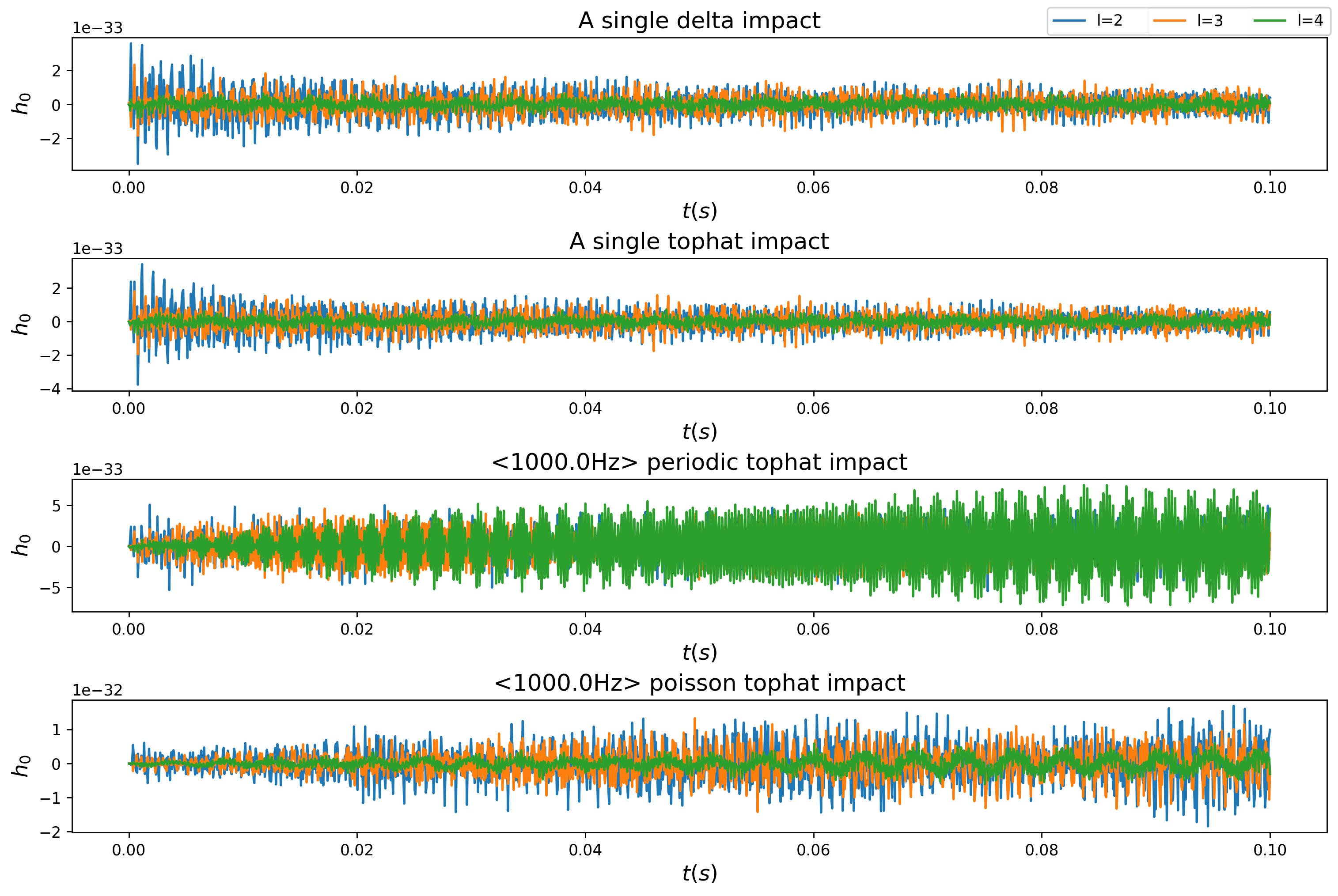}
        \caption{$n_{\text{poly}}=2, \Gamma_1=5/3$, $f$- and $p$-modes}
    \end{subfigure}
    \hfill
    \begin{subfigure}{\columnwidth}
        \includegraphics[width=\textwidth]{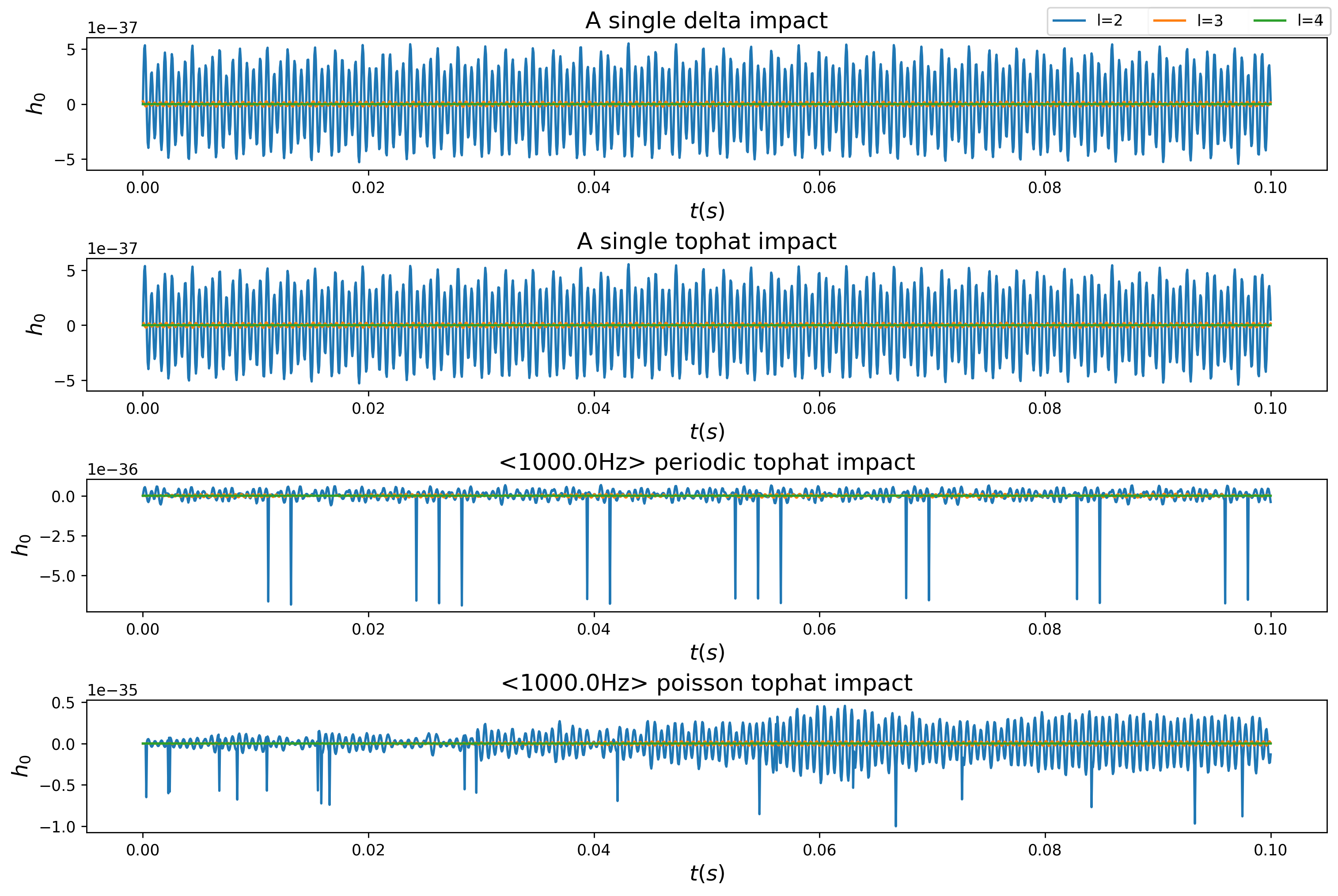}
        \caption{$n_{\text{poly}}=2, \Gamma_1=5/3$, $g$-modes}
    \end{subfigure}
    \caption{
    Gravitational waveforms \(h_0(t)\) emitted by \(l=2\) (blue curves), \(l=3\) (orange curves), and \(l=4\) (green curves) multipoles for three EOS:
    (a) \(n_{\text{poly}} = 1\), \(\Gamma_1 = 2\);
    (b) \(n_{\text{poly}}=1.5, \Gamma_1=5/3\); 
    (c) and (d) \(n_{\text{poly}}=2, \Gamma_1=5/3\).
    $f$-, $p$-, and $g$-modes are overlaid in quadrants (a) and (b) and are plotted separately in quadrants (c) and (d).
    Each quadrant (a)--(d) contains four subpanels corresponding to the four types of impact defined in Section~\ref{subsec: 5-Types_of_impact}: 
    single delta impact, single top-hat impact, periodic top hat, and Poisson top hat (ordered from top to bottom). 
    The waveforms are sampled at \(\SI{16384}{Hz}\) to match the data acquisition frequency in LVK detectors \citep{AbbottEtAl2021a}. 
    Accretion parameters: \(\Dot{M}=\SI{e-8}{\Msun/yr}, f_{\text{acc}} = \SI{1}{kHz}, \vb{v} = -c\hat{\vb{r}}\). 
    Stellar parameters: \(M=\SI{1.4}{\Msun}\), \(R=\SI{10}{km}\).}
    \label{fig:waveforml234}
\end{figure*}

If the accreting gas is guided magnetically onto the magnetic polar cap, then the impact footprint is approximately axisymmetric, and modes with \(m=0\) dominate.\footnote{
The reader is referred to \citet{KulkarniRomanova2008} for situations where the accretion footprint is not axisymmetric.
}
In what follows, we present results for the special case of even multipoles (where \(l\geq 2\) is an even integer) to illustrate by way of example some of the diverse waveforms that are possible. With \(\sigma_{nl}\tau_{nl}\gg1\) and writing \(h_{jk}^{\text{TT}} (t) = h_0(t) T^{\text{E2},l0}_{jk}\), we obtain 
\begin{align}
    h_0(t) 
    \propto& \, 
    \frac{G}{c^{l+2} T d} 
    \sum_{s=1}^{N(t)} \sum_{n\leq n_{\text{c}}} 
    \qty(-1)^{l/2} \vb{p}\vdot \boldsymbol{\xi}^{*(nl0)}(\vb{x}_\text{f}) Q_{nl} \sigma_{nl}^{l-2} 
    \nonumber \\ &\times 
    \Bigg\{
        \exp[-(t-t_{\text{s}}-T) / \tau_{nl}] \cos[\sigma_{nl}(t-t_s-T)] H(t-t_s-T) 
    \nonumber \\ 
        &- \exp[-(t-t_{\text{s}}) / \tau_{nl}] \cos[\sigma_{nl}(t-t_s)] H(t-t_s)
    \Bigg\} \,.
\end{align}
Interestingly, for \(l=2\), \(\sigma_{nl}\) appears explicitly in the phase but not the amplitude of \(h_0(t)\), for a top-hat impact.
However, the amplitude does depend on \(\sigma_{n2}\) implicitly. One can see this by replacing the sum over \(s\) with its continuous analog (i.e. an integral), whereupon one obtains \(h_0 \propto \sum_{n} \sigma_{n2}^{-1} \times (\text{terms depending on }n)\) approximately.

\subsection{Types of impact}
\label{subsec: 5-Types_of_impact}

In this subsection, we evaluate \(h_0(t)\) for an illustrative selection of the impact models in Section~\ref{sec: 4-Accretion_excitation}: a single delta-function in time, a single top hat, a sequence of top hats which is periodic, and a sequence of top hats which is Poisson (uniformly) distributed in time, following \eqref{eqn: shot_process}.

Figure~\ref{fig:waveforml234} shows the dominant \(m=0\) quadrupolar gravitational waveforms from accretion-induced oscillations. 
Each quadrant (a)--(d) of the figure contains four subpanels, which display \(h_0(t)\) versus $t$ for the four kinds of impact defined in the previous paragraph, ordered from top to bottom. The top two quadrants (a) and (b) correspond to relatively stiff EOS, with \(n_{\text{poly}} \leq 1.5\). The bottom two quadrants (c) and (d) correspond to a softer EOS (\(n_{\text{poly}}=2\)), with the $f$- and $p$-modes (bottom left quadrant) and $g$-modes (bottom right quadrant) plotted separately for clarity. 
Gravitational radiation reaction is included in the plots; see Section~\ref{subsec: 5-Damping_timescale} for details about the damping timescale. 
The waveforms produced by a top-hat impact and a delta-function impact are similar in shape and amplitude. There is a burst of activity lasting \(\lesssim 10^{-3}\)s after the impact, during which an amplitude-modulated ``packet'' of shorter-period modes is excited. The packet evolves into a longer-period oscillation for \(t \gtrsim 10^{-3}\)s. 
The similarity between delta-function and top-hat impacts arises, because both are shorter than the periods of modes that dominate the GW emission. 
The strain from periodic impacts is only \(\sim 2\) times greater than the strain produced by one impact, because all modes have frequencies different from \(f_{\text{acc}}\) and therefore are not excited resonantly. 
Periodic impacts form amplitude-modulated packets, whose envelopes oscillate with period \(\sim 10^{-2}\)s, whereas the waveform produced by Poisson impacts is shaped irregularly.
The strain from Poisson impacts is $2$--$4$ times greater than from periodic impacts due to the randomness in \(t_{\text{s}}\), which disrupts the cancellation involved in non-resonant periodic excitation.

The GW signal from $g$-modes, plotted separately for clarity in quadrant (d), is considerably weaker (by about four orders of magnitude) than from $f$- and $p$-modes for three reasons. First, $g$-modes have smaller \(Q_{nl}\) because of more complete cancellation in the oscillatory integrand in \eqref{eqn: Q_expanded}. Second, the accreting material is assumed to strike the surface radially, and $g$-modes have smaller \(\hat{\bf r}\vdot \boldsymbol{\xi}^{*(\alpha)}(\vb{x}_{\rm f})\) than $p$-modes. Third, the GW signal is proportional to the time derivatives of \(I_{lm}\), and $g$-modes oscillate slower than $f$- and $p$-modes. 

Polytropes with stiffer EOS lead to smaller mode amplitudes for the same impact. 
The overlap integral \(Q_{nl}\) is also smaller (except for $f$-modes), because the wave strain contribution from expanding and contracting mass shells cancel more completely in a less centrally condensed star.
Hence, polytropes with smaller \(n_{\text{poly}}\) emit weaker GW, as confirmed by Figure~\ref{fig:waveforml234}. We find \(h_0 \approx 1\times 10^{-33}\), \(5\times10^{-33}\), and \(2\times 10^{-32}\) in quadrants (a), (b), and (c) respectively.

Multipole order plays a significant role in setting the shape and amplitude of the emitted waveform. The discussion in the previous paragraphs applies principally to the \(l=2\) multipole, which is drawn as a blue curve throughout Figure~\ref{fig:waveforml234}. We overlay the \(l=3\) and \(l=4\) signals as orange and green curves respectively. 
The ``packets'' produced by periodic impacts have similar periods for \(l=3\) and \(l=4\). 
The single-impact waveforms show that \(l=2\) dominates usually. However, the quadrupole GWs emitted by $f$- and $p$-modes are damped until they reach similar magnitude to higher orders within \(\approx 3\tau_{\alpha}\) for $^{2}f$-modes (\(\sim \SI{0.02}{ms}\)).
The amplitude of the strain emitted by $f$- and $p$-modes from \(l=3\) and \(l=4\) multipoles is an order of magnitude smaller than from \(l=2\) for \(n_{\text{poly}}=1.0\) and of the same order for \(n_{\text{poly}}=1.5\) and \(2\).
The quadrupole GWs emitted by $g$-modes always dominate over higher-order multipoles.

\subsection{Dominant modes}
\label{subsec: 5-Dominant_modes}
In this subsection, we explore the energy spectra of GW signals generated by multiple Poisson top-hat impacts with different \(T\), in order to compare with the simulated GW signals from a single, infalling, quadrupolar mass shell in \citet{Nagar_etal2004}.

\begin{figure}
    \centering
    \begin{subfigure}{\columnwidth}
        \includegraphics[width=\textwidth]{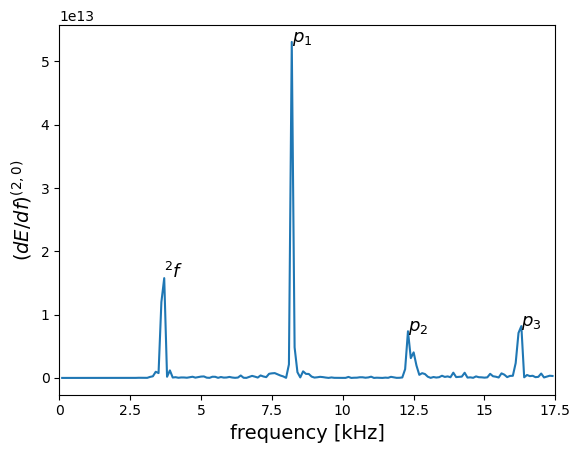}
        \caption{}
        \label{subfiga:compare_with_Nagar2004}
    \end{subfigure}
    \vfill
    \begin{subfigure}{\columnwidth}
        \includegraphics[width=\textwidth]{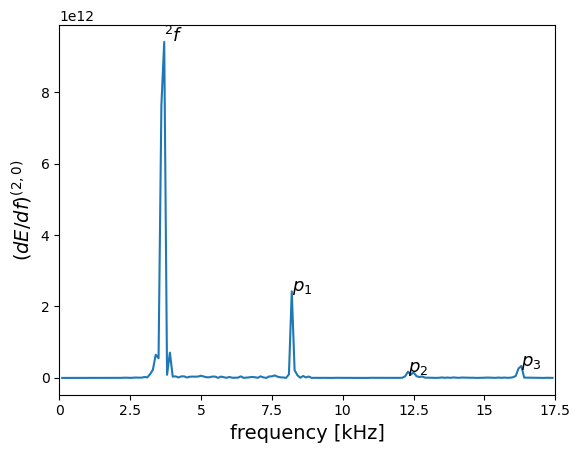}
        \caption{}
        \label{subfigb:compare_with_Nagar2004}
    \end{subfigure}
    \vfill
    \begin{subfigure}{\columnwidth}
        \includegraphics[width=\textwidth]{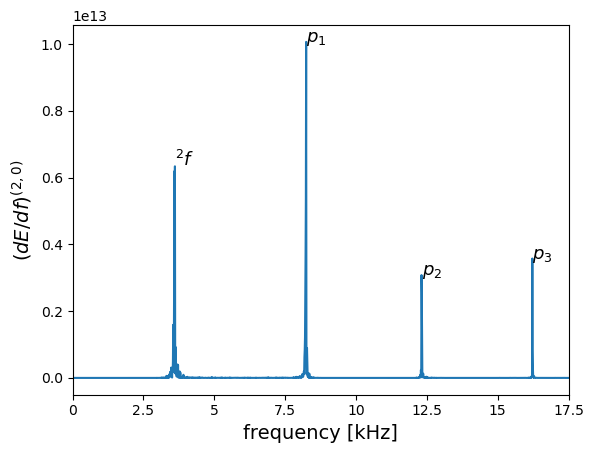}
        \caption{}
        \label{subfigc:compare_with_Nagar2004}
    \end{subfigure}
    \caption{Energy spectral density \(dE^{(2,0)}/df\) as a function of Fourier frequency $f$ for GW signals produced by Poisson top-hat impacts. (a) Spectral window $10$ms, $T=10^{-5}$s. (b) Spectral window $10$ms, $T=10^{-4}$s. (c) Spectral window $100$ms, $T=10^{-4}$s.
    The sampling frequency is chosen to be $35$kHz to avoid aliasing.}
    \label{fig:compare_with_Nagar2004}
\end{figure}

Figure~\ref{fig:compare_with_Nagar2004} displays the energy spectral density \(dE^{(l,m)}/df\) of the GW signals emitted by oscillations excited by a sequence of Poisson top-hat impacts for \((l,m)=(2,0)\). 
The time span used to calculate \(dE^{(l,m)}/df\) by Fourier transforming \(h_0(t)\) is termed the spectral window.\footnote{For the sake of definiteness, the spectral window begins at \(t=0\) in this paper.}
In panels (a)--(c), we present, from top to bottom, \(dE^{(2,0)}/df\) for (a) \(T=10^{-5}\)s with a spectral window of $10$ms, (b) \(T=10^{-4}\)s with a spectral window of $10$ms, and (c) \(T=10^{-4}\)s with a spectral window of $100$ms respectively.
The spectral window used by \citet{Nagar_etal2004} is \(\sim 10\)ms.
Figure~\ref{subfiga:compare_with_Nagar2004} shows that $p_1$ contains most of the energy during the initial \(10\) ms for \(T=10^{-5}\)s, whereas the $f$-mode dominates in the simulations by \citet{Nagar_etal2004}. 
The difference arises because $p$-modes with \(n\geq2\) are suppressed for \(T=10^{-5}\)s through top-hat averaging, except for the \(p_1\) mode. 
In Figure~\ref{subfigb:compare_with_Nagar2004}, the $f$-mode emerges as the highest peak, because the \(p_1\) mode is suppressed for \(T=10^{-4}\)s, which is of the same order of magnitude as \(T\) in \citet{Nagar_etal2004} (see the text above equation (43) in the latter reference). 

Another essential distinction between this paper and \citet{Nagar_etal2004} is that this paper focuses on long-lasting accretion, while \citet{Nagar_etal2004} applies to transient core-collapse scenarios. 
In Figure~\ref{subfigc:compare_with_Nagar2004}, we plot \(dE^{(2,0)}/df\) for a longer spectral window of \(100\) ms. $p_1$ re-emerges with the highest peak. 
Figure~\ref{subfigb:compare_with_Nagar2004} and \ref{subfigc:compare_with_Nagar2004} imply that more energy accumulates in higher-$n$ modes, as the spectral window lengthens (see Section~\ref{subsec: 6-Temporal_autocorrelation_function}).

\subsection{Damping timescale}
\label{subsec: 5-Damping_timescale}
A variety of damping mechanisms exist in realistic neutron stars, including but not limited to gravitational radiation reaction, neutrino emission, nonadiabaticity, internal friction, and electromagnetic radiation. 
Nonadiabaticity and viscosity are insignificant as explained in Section~\ref{sec: 3-Linear_adiabatic_pulsation_theory}. Electromagnetic damping and neutrino damping can be important for certain modes \citep{McDermott_etal1988}. 
However, gravitational radiation reaction dominates the damping for the $f$-,$p$-, and $g$-modes we study in this paper. 

We estimate the damping timescale a posteriori via \(\tau_{nl} = 2E_{\alpha}/L_{\text{gw}}\), where \(L_{\text{gw}}\) is the power of the gravitational radiation, a standard approach \citep{Thorne1969b, BalbinskiSchutz1982, McDermott_etal1988, OwenEtAl1998}. 
In our notation, the \(e\)-folding timescale reads 
\begin{align}
    \tau_{nl} = \frac{l(l-1)[(2l+1)!!]^2}{2\pi (l+1)(l+2)} 
                \qty(\frac{c}{R_{*} \sigma_{nl}})^{2l+1} 
                \frac{J \, R_{*} \sigma_{nl}}{G M_{*}^2 Q_{nl}^2} 
                \,.
    \label{eqn: damping_time_GW}
\end{align}
The damping modification of \(\sigma_{nl}\) is of order \(\order{\sigma_{nl}^{-2} \tau_{nL}^{-2}}\) and is therefore insignificant for \(\sigma_{nl}\tau_{nl} \gg1\), as typically occurs. Numerical results for the damping timescale are given in Table~\ref{tab:M1.4_R1.0_L2}. \(\tau_{nl} \propto Q_{nl}^{-2}\) grows exponentially with \(n\) because \(Q_{nl}\) decreases exponentially with \(n\) as demonstrated in Appendix~\ref{app:overlap_integral}. Therefore, gravitational radiation reaction does not damp high-$n$ modes efficiently.

\section{Detectability}
\label{sec: 6-Detectability}
In this section, we calculate the temporal autocorrelation function of the stochastic GW strain calculated in Section~\ref{sec: 5-Gravitational_waveform} with the aid of the analytic formulation in Appendix~\ref{app:statistics_for_the_shot_process}. 
The autocorrelation function is important for two reasons in this paper:
\begin{enumerate*}[(i)]
    \item when evaluated at zero time lag, it returns the mean square amplitude of the GW signal, a key quantity when assessing detectability; and
    \item when Fourier transformed, it returns the power spectral density of the GW signal, which contains important information about the source physics, e.g. the modes excited by accretion, and the statistics of the accreted clumps.
\end{enumerate*}
In Section~\ref{subsec: 6-Temporal_autocorrelation_function}, we calculate the temporal autocorrelation function of \(h_0(t)\).
We focus on the autocorrelation of modes with themselves and de-emphasise the cross-correlation between different modes, which dephase due to frequency mismatch over time lags that are relevant to GW observations. 
The detectability of the GW signal is investigated in both time and frequency domains, via the root-mean-square wave strain and the amplitude spectral density (ASD) in Sections~\ref{subsec: 6-Root-mean-square wave strain} and \ref{subsec: 6-Amplitude_spectral_density} respectively.

\subsection{Temporal autocorrelation function}
\label{subsec: 6-Temporal_autocorrelation_function}
We define the autocorrelation function as
\begin{align}
    C(t, t') = \expval{h_0(t) h_0^{*}(t')} \,,
    \label{eqn: autocorrelation_h0_definition}
\end{align}
where angular brackets denote an ensemble average.
For the sake of simplicity, we focus on the dominant multipole \((l,m)=(2,0)\).
For a stationary process, we have \(C(t,t')=C(\abs{\zeta})\), where \(\zeta=t'-t\) is the time lag between measurements. 
We find
\begin{align}
    C(\zeta) 
    &= \frac{256\pi^2 G^2 M_{*}^2 R_{*}^4}
        {75 c^{8} d^2} 
        \sum_{n, n'} Q_{n}Q_{n'} 
        \expval{\dv[2]{t} A_{n}(t) 
        \frac{d^2}{dt^{'2}} A_{n'}(t')} ,
    \label{eqn:autocorrelation_h0_definition2}
\end{align}
with \(\delta\vb{r}(t) = \sum_{\alpha} A_{\alpha}(t) \boldsymbol{\xi}^{(\alpha)}\). 
The indices \((l,m)=(2,0)\) are suppressed for convenience in \(A_{n}, \sigma_{n}\), and \(\tau_{n}\) for the remainder of this subsection. 
The statistical properties of \(A_{\alpha}(t)\)-like random variables, including their autocorrelation, are derived in Appendix~\ref{app:statistics_for_the_shot_process}. From Appendix~\ref{app:statistics_for_the_shot_process}, we have 
\begin{align}
    &\expval{\dv[2]{t} A_{n}(t) \frac{d^2}{dt^{'2}} A_{n'}(t')} 
    \nonumber \\
    =&\,f_{\text{acc}}^2 
        \int_{-\infty}^t dt_{\text{s}} \int_{-\infty}^{t'} dt'_{\text{s}} 
        \expval{\dv[2]{t} A_{n}(t; t_{\text{s}})}
        \expval{\frac{d^2}{dt^{'2}} A_{n'}(t'; t'_{\text{s}})}
        \nonumber \\ &+ 
        f_{\text{acc}} \int_{-\infty}^{\min\{t,t'\}} dt_{\text{s}} 
        \expval{\dv[2]{t} A_{n}(t; t_{\text{s}}) \frac{d^2}{dt^{'2}} A_{n'}(t'; t_{\text{s}})} \,.
    \label{eqn:autocorrelation_amplitude}
\end{align} 
The double integral in the first line of \eqref{eqn:autocorrelation_amplitude} is zero as it involves the derivatives of constant mean amplitudes.
The second line in \eqref{eqn:autocorrelation_amplitude} is negligible, except when two modes have the same frequencies. Integrating the second line in \eqref{eqn:autocorrelation_amplitude} with \(n=n'\), we isolate the dominant term
\begin{align}
    \expval{\dv[2]{t} A_{n}(t) \frac{d^2}{dt^{'2}} A_{n}(t')} 
    =
    \frac{\abs{\vb{p}\vdot \boldsymbol{\xi}^{(n)}(\vb{x}_\text{f})}^2
    f_{\text{acc}} \tau_{n} 
    }{4 J^2 T^2 \sigma_{n}^4} 
    \dv[4]{\zeta} F(\zeta, T)
     \,,
    \label{eqn: autocorrelation_A_sameMode}
\end{align}
with
\begin{align}
    F(\zeta, T) 
    =&\, 2 \exp(-\abs{\zeta} / \tau_{n}) \cos{\sigma_{n} \abs{\zeta}} 
    \nonumber \\ 
    &- \exp[-(\abs{\zeta}+T) / \tau_{n}] \cos[\sigma_{n}(\abs{\zeta}+T)]
    \nonumber \\ 
    &- H(\abs{\zeta}-T) \exp[-(\abs{\zeta}-T) / \tau_{n}] \cos[\sigma_{n}(\abs{\zeta}-T)] 
    \nonumber \\ 
    &- [1-H(\abs{\zeta}-T)] \exp[(\abs{\zeta} -T) / \tau_{n}] \cos[\sigma_{n}(\abs{\zeta}-T)] 
    \,.
    \label{eqn: autocorrelation_A_F(zeta,T)}
\end{align}
Equation~\eqref{eqn: autocorrelation_A_sameMode} suggests that the amplitude autocorrelation increases with \(n\), because \(\tau_n\) increases with \(n\); that is, higher-order modes accumulate more energy from multiple impacts before being damped. 
The exponential factor \(\exp(-\abs{\zeta} / \tau_{n})\) in \eqref{eqn: autocorrelation_A_F(zeta,T)} implies that \(\expval{A_{n}(t) A_{n}(t')}\) decreases for \(\abs{\zeta} \gtrsim \tau_n\). The cross-correlation \(\expval{A_{n}(t) A_{n'}(t')}\) (for \(n<n'\), say) is bound by \(\expval{A_{n}(t) A_{n}(t')} < \expval{A_{n}(t) A_{n'}(t')} < \expval{A_{n'}(t) A_{n'}(t')}\) for \(\abs{\zeta} \gg \tau_n\).

Upon substituting \eqref{eqn: damping_time_GW} and \eqref{eqn: autocorrelation_A_sameMode} into \eqref{eqn:autocorrelation_h0_definition2} and neglecting the cross terms, \(C(\zeta)\) simplifies to 
\begin{align}
    C(\zeta) 
    =&\,
    7.8\times10^{-68} \gamma_{v}^2
    \qty(\frac{d}{\SI{1}{kpc}})^{-2} 
    \qty(\frac{M_{*}}{1.4M_{\odot}})^{-3}
    \qty(\frac{R_{*}}{10\si{km}})^{6}
    \nonumber \\ &\times
    \qty(\frac{\dot{M}}{10^{-8}M_{\odot}/\si{yr}})^2 
    \qty(\frac{f_{\text{acc}}}{\SI{1}{kHz}})^{-1}
    \qty(\frac{T}{10^{-5}\si{s}})^{-2} 
    \qty(\frac{\abs{\vb{v}}}{0.4c})^2
    \nonumber \\ &\times
    \sum_{n\leq n_{\text{c}}} 
    \abs{\frac{\hat{\vb{v}}\vdot \boldsymbol{\xi}^{(n)}(\vb{x}_\text{f})}{R_* \tilde{\sigma}_{n}^2}}^2
    F(\zeta, T) 
    \,,
    \label{eqn: autocorrelation_h0_scaling}
\end{align}
with \(\gamma_v = 1/\sqrt{1-(\abs{\bf v}/c)^2}\).
Interestingly, \(C(\zeta)\) does not depend on \(Q_{nl}\), because \(\tau_{nl}\) cancels \(Q_{nl}\) and we have \(d^{2l}/d\zeta^{2l} \propto \sigma_{nl}^{2l}\).
The cancellation occurs when GW damping dominates; a different scaling applies for other damping mechanisms, to be studied in future work.
We redefine the mode cutoff \(n_\text{c}\) to incorporate this \(Q_{nl }\)-independence, ensuring \(\tau_{nl} \lesssim \Msun/\dot{M}\) (e.g. \(n_\text{c} \approx 15\) for \(n_{\text{poly}}=1.0\)). Modes with \(n > n_c\) generate negligible \(h_0\) because the accumulation of energy from multiple impacts cannot compensate for the exponentially decaying \(Q_{nl}\).
To understand this concept mathematically, consider the rate of change of the energy \(E_{\alpha}\) in mode \(\alpha\), viz. 
\begin{align}
    \dv*{E_{\alpha}}{t} = P^{\text{in}}_{\alpha} - E_{\alpha}/\tau_{\alpha} , \tag{\textasteriskcentered} \label{eqn:dE/dt}
\end{align} where \(P^{\text{in}}_{\alpha}\) is the input power. 
Consider a higher-order mode A with \(\tau_{\rm A} > t_0\) and a lower-order mode B with \(\tau_{\rm B} < t_0\), where \(t_0\) is the lifetime of accretion episode. 
The ratio of the emitted GW power at \(t = t_0\) is then \([E_{\rm A}(t_0) / \tau_{\rm A}] [E_{\rm B}(t_0) / \tau_{\rm B}] \approx (P^{\text{in}}_{\rm A} / P^{\text{in}}_{\rm B}) (t_0 / \tau_{\rm A})\) from solving \eqref{eqn:dE/dt}.
Since the damping time-scale increases exponentially with increasing mode order \(n\), and \(P^{\rm in}\) does not increase as fast, we have \((P^{\text{in}}_{\rm A} / P^{\text{in}}_{\rm B}) (t_0 / \tau_{\rm A}) \ll 1\).
When other damping mechanisms (e.g., neutrino emission or viscosity) are considered, the power of the signal is scaled by \(\sim \tau_{\text{total}} / \tau_{\text{GW}}\), with \(1/\tau_{\text{total}} = 1/\tau_{\text{GW}} + 1/\tau_{\text{other}}\), where \(\tau_{\text{GW}}\) and \(\tau_{\text{other}}\) are the damping time-scales via gravitational radiation and another mechanism, respectively. Therefore, modes with \(\tau_{\text{other}} \ll \tau_{\text{GW}} < t_0\) also contribute negligibly to the long-term signal.
Unlike \eqref{eqn: displacement_tophat_gtrT}, the factor \(|\boldsymbol{\xi}^{(n)}| / (R_* \tilde{\sigma}_n^2)\) in \eqref{eqn: autocorrelation_h0_scaling} increases with \(n\) for a relatively soft EOS (e.g. \(n_{\text{poly}}=1.5\) or $2.0$), but decreases with increasing \(n\) for a relatively stiff EOS (e.g \(n_{\text{poly}}=1.0\)).

\subsection{Root-mean-square wave strain}
\label{subsec: 6-Root-mean-square wave strain}
As a first pass, we assess the detectability of the GW signal in the time domain. We calculate the root-mean-square characteristic wave strain \(h_{\text{rms}}\) in terms of the temporal autocorrelation function at zero lag, viz. \(h_{\text{rms}} = C(\zeta=0)^{1/2}\) and hence
\begin{align}
    h_{\text{rms}} =& \, 2.8\times10^{-34} \gamma_{v}
    \qty(\frac{d}{\SI{1}{kpc}})^{-1} 
    \qty(\frac{M_{*}}{1.4M_{\odot}})^{-3/2}
    \qty(\frac{R_{*}}{10\si{km}})^{3}
    \nonumber \\ &\times
    \qty(\frac{\dot{M}}{10^{-8}M_{\odot}/\si{yr}})
    \qty(\frac{f_{\text{acc}}}{\SI{1}{kHz}})^{-1/2}
    \qty(\frac{T}{10^{-5}\si{s}})^{-1} 
    \qty(\frac{\abs{\vb{v}}}{0.4c})
    \nonumber \\ &\times
    \qty{ \sum_{n,l} 
    \abs{\frac{\hat{\vb{v}}\vdot \boldsymbol{\xi}^{(nl)}(\vb{x}_{\rm f})}{R_* \tilde{\sigma}_{nl}^2}}^2 
    \qty[2 - 2\exp(-T/\tau_{nl}) \cos(\sigma_{nl} T)]
    }^{1/2}
    .
    \label{eqn: h_rms}
\end{align}
Equation~\eqref{eqn: h_rms} follows immediately from \eqref{eqn: autocorrelation_h0_scaling}, except that the sum in the last line is over all \(l\geq2\), not just \(l=2\). 
Note that equation~\eqref{eqn: h_rms} implies \(h_\text{rms} \propto T^{-1} \rightarrow \infty\) if one naively takes the limit \(T\rightarrow 0\). The divergence occurs, because one has \(\mathcal{T} \propto T^{-1}\) nominally in Equation~\eqref{eqn: shot_process} for a top-hat impact. 
The divergence disappears, if one directly computes the ordered limit \(\lim_{\zeta \rightarrow 0} \lim_{T\rightarrow 0} F(\zeta, T)\) in \eqref{eqn: autocorrelation_A_F(zeta,T)}.
Note also that \(M_*, R_*\) and \(\tilde\sigma_{\alpha}\) are related through the EOS. For example, one must simultaneously scale \(M_*\) and \(R_*\) according to the EOS, in order to leave \(\tilde\sigma_{\alpha}\) unchanged.

Equation~\eqref{eqn: h_rms} implies that the GW signal emitted by accretion-driven stellar oscillations is too weak to be detected by the current generation of LVK interferometers \citep{WattsEtAl2008, HaskellEtAl2015}. 
The value of \(h_{\text{rms}}\) in \eqref{eqn: h_rms} is lower than the upper bounds \(h_{0}^{95\%} \sim 10^{-26}\) (at 95\% confidence) inferred from recent searches of continuous waves from accreting systems \citep{MiddletonEtAl2020a, ZhangEtAl2021, CovasEtAl2022, LIGO2022_20AccretingPulsar, LIGO2022a_allsky}. 

Equation~\eqref{eqn: h_rms} is consistent with an order-of-magnitude estimate of \(h_{\text{rms}}\) following from the energy conservation argument \(L_{\text{gw}}\sim \eta_{\text{gw}}G\dot{M}M_*/R_* \approx \eta_{\text{gw}}\dot{M}\abs{\vb{v}}^2/2\), where \(\eta_{\text{gw}}\) is the energy conversion efficiency. If we take \(\eta_{\text{gw}} = m/M_* = \dot{M}/(f_{\text{acc}}M_*)\), as for an inelastic collision of two point masses $m$ and \(M_{*}\) in the extreme mass-ratio limit, energy conservation implies 
\begin{align}
    h_{\text{rms}} \sim& \, 
    \qty(\frac{4 G \dot{M} \abs{\vb{v}}^2 \eta_{\text{gw}}}{d^2 c^3 f_{\text{acc}}^2})^{1/2}
    \label{eqn:h_rms_with_etaGW}
    \\
    =& \, 1.47\times10^{-34} 
    \qty(\frac{M_*}{1.4\Msun})^{-1/2}
    \qty(\frac{d}{\si{kpc}})^{-1} 
    \qty(\frac{\abs{\vb{v}}}{0.4c})
    \nonumber \\ &\times
    \qty(\frac{\dot{M}}{10^{-8}M_{\odot}/\si{yr}})
    \qty(\frac{f_{\text{acc}}}{\si{kHz}})^{-3/2}
    \,,
    \label{eqn:h_rms_inelastic}
\end{align}
which is comparable in value and scaling to \eqref{eqn: h_rms} and implies \(h_{\text{rms}}\lesssim 10^{-25}\) for \(\eta_{\text{gw}} \leq 1\). 
The additional factors in \eqref{eqn: h_rms} carry information about the impact (e.g. \(T\)) and the EOS (e.g. \(\sigma_{\alpha}\), which also affects the scalings with \(M_*\) and \(R_*\)). 
For example, when taking \(\hat{\bf p} = \hat{\bf r}\), we find \([\sum_{l=2}^{4} \sum_{m} \sum_{\abs{n}\leq n_c} |\hat{\bf p} \vdot \boldsymbol{\xi}^{(\alpha)}(\vb{x}_{\rm f}) / R_* \tilde{\sigma}_{\alpha}^2|^2 F(\zeta=0,T)]^{1/2} \approx 3.5, 18\), and \(1.0\times10^2\) for \(n_\text{poly}=1.0, 1.5\), and 2 respectively. 
By contrast, when taking \(\hat{\bf p} = \hat{\phi}\), we find \([\sum_{l=2}^{4} \sum_{m} \sum_{\abs{n}\leq n_c} |\hat{\bf p} \vdot \boldsymbol{\xi}^{(\alpha)}(\vb{x}_{\rm f}) / R_* \tilde{\sigma}_{\alpha}^2|^2 F(\zeta=0,T)]^{1/2} \approx 0.44, 0.69\), and \(1.3\times10^2\) for \(n_\text{poly}=1.0, 1.5\), and 2 respectively.
The approximation \eqref{eqn:h_rms_inelastic} converges to \eqref{eqn: h_rms}, as \(n_{\text{poly}}\) decreases, because stars with stiffer EOS behave more like rigid bodies. 

The accretion torque exerted by accreted matter is \(\dot{J}_{\text{acc}} = \dot{M}(GM_* R_*)^{1/2} = L_{\text{acc}} (G M_* / R_*^3)^{-1/2}\). On the other hand, the torque of the gravitational radiation emitted from the \((l, m)\)-th multipole can be estimated via \(\dot{J}_{\text{GW}} = \dot{E}_{\text{GW}} (\sigma/m)^{-1} \sim m (h/10^{-25})^2 (\dot{M}/10^{-8}\si{M_{\odot}/yr})^{-1} \dot{J}_{\text{acc}}\) \citep{FriedmanSchutz1978}, where \(m\) is the order of the spherical harmonic \(Y_{l}^{m}\).
We have \(h_{\rm rms} \lesssim 10^{-32} \ll 10^{-25}\) for \(\dot{M} = 10^{-8}\si{M_{\odot} yr^{-1}}\), so the gravitational radiation torque in the mechanical excitation scenario is negligible, compared to the accretion torque.

\subsection{ASD}
\label{subsec: 6-Amplitude_spectral_density}
A complementary approach to assessing the detectability of the GW signal is to compute its ASD, \(S_h(f)^{1/2}\), in the frequency domain and compare it with the ASD of the LVK interferometers. This is important, as some of the signal power \(\propto h_{\text{rms}}^2\) falls outside the LVK observing band and cannot be detected. Conversely, predicting the ASD theoretically becomes important, once a signal is detected, because the locations of the spectral peaks contain important information about the stiffness of the EOS. For example, high-order modes are prominent, if the EOS is relatively soft. 

\begin{figure}
    \centering
    \includegraphics[width=\columnwidth]{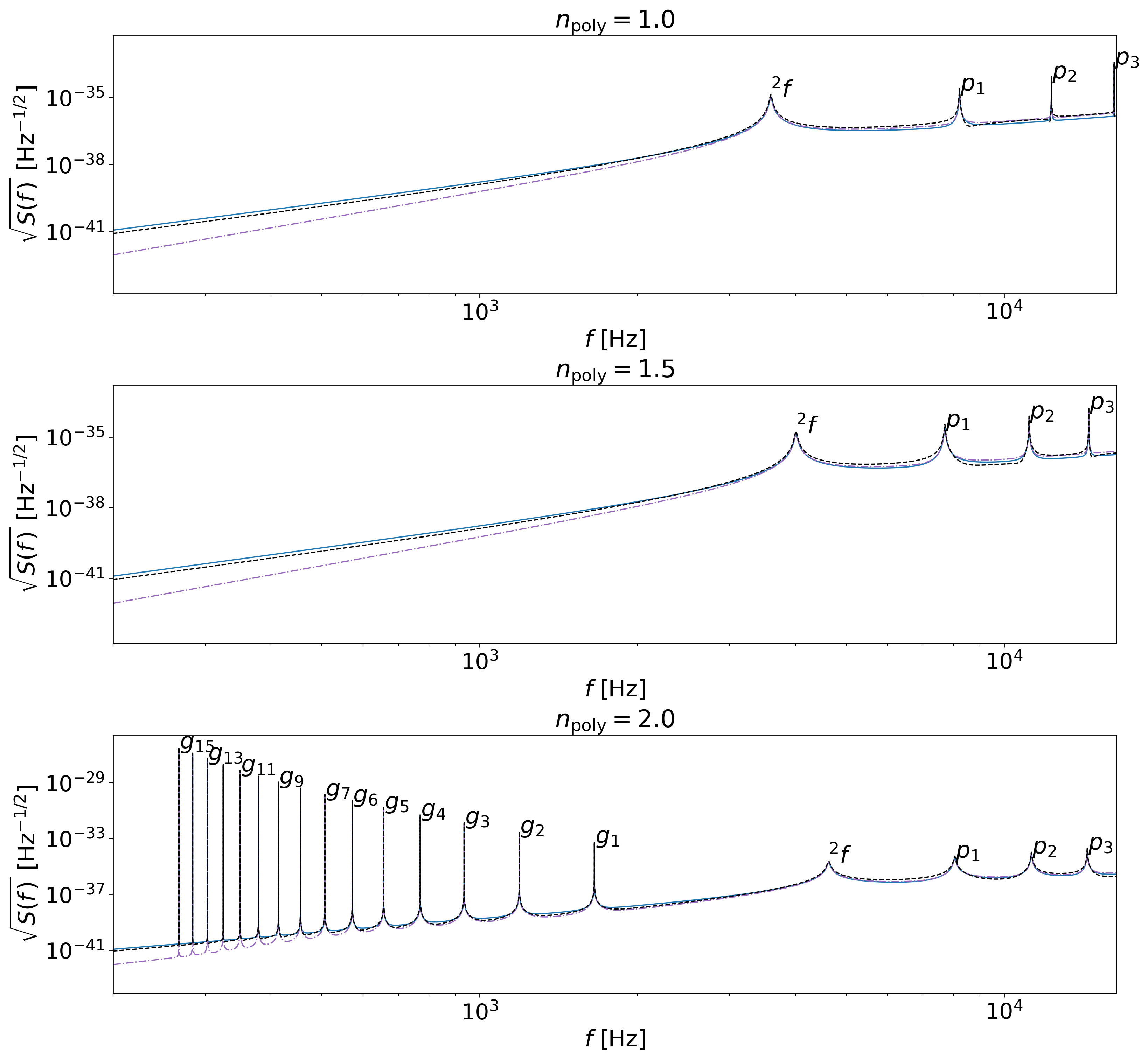}
    \caption{ASD \(S_{h}(f)^{1/2}\) (units: Hz$^{-1/2}$) versus frequency (units: Hz) within the upper LVK observing band:
    approximate ASD for \(l=2\) including only the dominant terms in \(C(\zeta)\), i.e. equation~\eqref{eqn: autocorrelation_h0_scaling} (solid blue curve), approximate ASD excluding cross-correlations (dash-dot purple curve) and exact ASD (dashed black curve) including all terms in \(C(\zeta)\) and cross-correlations.
    Each subpanel corresponds to a different EOS: \(n_{\text{poly}}=1.0\) (top subpanel), 1.5 (middle subpanel), and 2.0 (bottom subpanel). Modes are labelled as in the text, e.g. \(^{l}p_n\). Every second $g$-mode with \(n\geq7\) is not labelled for readability.
    }
    \label{fig:ASD_h0}
\end{figure}

\begin{figure}
    \centering
    \includegraphics[width=\columnwidth]{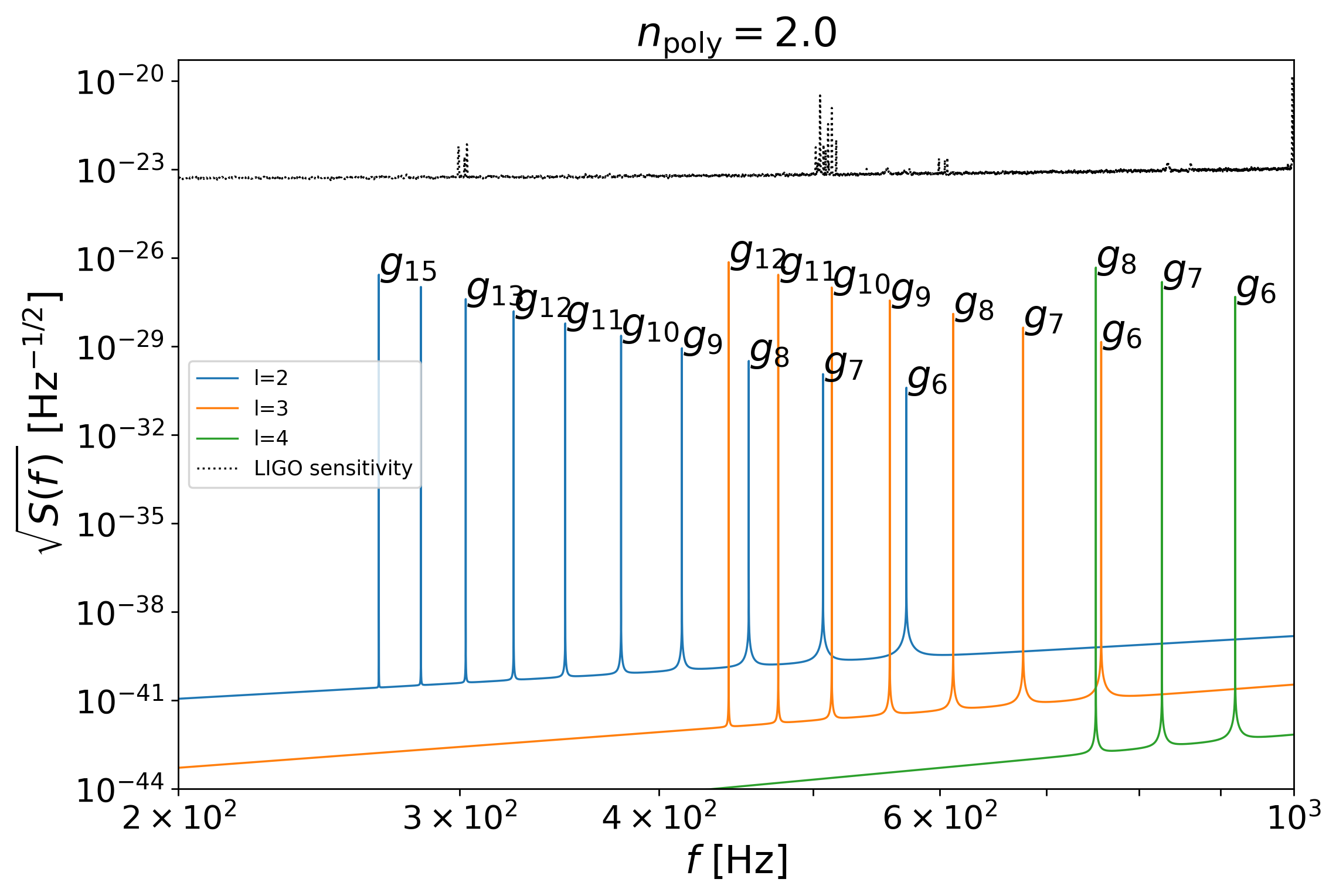}
    \caption{Close-up view of the ASD for \(l=2\) (blue curve), \(l=3\) (orange curve), and \(l=4\) (green curve) plotted with the sensitivity curve (dotted black curve) achieved during the third LVK observing run.
    The ASD is calculated using the leading-order approximation [i.e. equation~\eqref{eqn: autocorrelation_h0_scaling}], as for the blue curve in Figure~\ref{fig:ASD_h0}. 
    Physical parameters: \(n_{\text{poly}}=2.0\), \(M_{*}=1.4\Msun\), \(R_{*}=\SI{10}{km}\), \(\dot{M}=\SI{e-8}{\Msun \per yr}\), \(f_{\text{acc}}=\SI{1}{kHz}\).}
    \label{fig:ASD_near_LIGO_curve}
\end{figure}

The one-sided ASD is related to the Fourier transform of \(C(\zeta)\) by
\begin{align}
    S_h(f)^{1/2} = \qty[2 \int_{-\infty}^{\infty} d\zeta \, C(\zeta) e^{i 2\pi f\zeta}]^{1/2} \,,
\end{align}
where \(f>0\) is the Fourier frequency in Hz. Examples of the ASD are plotted in Figure~\ref{fig:ASD_h0} for three EOS with \(n_{\text{poly}} = 1.0, 1.5\), and 2.0 (top to bottom subpanels), and modes with \(l=2\) and \(n\leq n_c\) in the upper LVK observing band. 
The peaks reach \(S_{h}(f)^{1/2} \sim \SI{e-34}{Hz}^{-1/2}\) for $f$- and $p$-modes, and \(\SI{e-32}{Hz}^{-1/2} \lesssim S_{h}(f)^{1/2} \lesssim \SI{e-25}{Hz}^{-1/2}\) for $g$-modes. 
For \(n_{\text{poly}}=1.0\) and 1.5, only the $f$-mode falls within the calibrated observing band\footnote{\url{https://dcc.ligo.org/public/0177/T2100313/003/README}}
(\(\SI{10}{Hz} \lesssim f \lesssim \SI{5}{kHz}\)), and there are no $g$-modes (\(\Gamma_1 = \Gamma\)). 
For low-order modes (\(n\lesssim5\)), the predicted ASD drops below the sensitivity curve of the current generation of long-baseline interferometers, e.g. the LVK detectors \citep{CahillaneMansell2022}; see Figure 4 in the latter reference. 
However, the ASD increases with mode order in general, mainly because \(\tau_{nl}\) increases with mode order.
Moreover, $g$-modes generally attain larger \(S_h(f)^{1/2}\) than $f$- and $p$-modes because they have higher \(\tau_{nl}\) and lower \(\sigma_{nl}\).
The peaks in Figure~\ref{fig:ASD_h0} grow narrower, as $n$ increases.

Figure~\ref{fig:ASD_h0} demonstrates how various approximations, made for computational convenience, affect the ASD.
Each panel displays three curves for quadrupolar GWs: the exact ASD, that includes all terms and cross-correlations in \(C(\zeta)\) (dashed black curve); the ASD that excludes cross-correlations in \(C(\zeta)\) (dash-dot purple curve); and the ASD that includes only the dominant terms of \(C(\zeta)\), i.e. equation~\eqref{eqn: autocorrelation_h0_scaling} (solid blue curves).
We see that the solid blue curves align well with the dash-dot purple curves, except for \(f \lesssim \SI{1}{kHz}\).
The cross-correlation terms also do not modify \(S_h(f)^{1/2}\) much near the peaks (\(f \approx \sigma_\alpha\)), which implies that \eqref{eqn: autocorrelation_h0_scaling} is a good approximation. 
They do modify \(S_h(f)^{1/2}\) between the $p$-mode peaks, e.g. for \(\SI{5e3}{Hz} \lesssim f \lesssim \SI{e4}{Hz}\) in the middle panel of Figure~\ref{fig:ASD_h0}. 
This occurs because we have \(\expval{A_{n}(t) A_{n'}(t')} > \expval{A_{n}(t) A_{n}(t')}\) for \(n' \neq n\) and \(\abs{t'-t}\gg \tau_{n}\) (see Section~\ref{subsec: 6-Temporal_autocorrelation_function}).

Figure~\ref{fig:ASD_near_LIGO_curve} presents a magnified comparison between the sensitivity curve of a typical LVK detector during the third observing run and the ASD of higher-order (\(n\geq6\)) $g$-modes for \(l=2,3\), and 4, and \(n_{\text{poly}}=2.0\). 
The peaks of \(S_h(f)^{1/2}\) approach the sensitivity curve, as \(n\leq n_c\) increases, but are bounded by \(S_h(f)^{1/2}\lesssim 10^{-26}\si{Hz}^{-1/2}\).
However, the peaks are narrow, so their signal-to-noise ratio (SNR) \(\propto \int df\, S_h(f)^{1/2}\) is low, and they remain difficult to detect, in keeping with Section~\ref{subsec: 6-Root-mean-square wave strain}.
A rigorous calculation of the SNR requires information about the specific detection algorithm. 
It goes beyond the scope of this paper and will be tackled in future work.

\section{Conclusion}
\label{sec: 7-Conclusion}
In this paper, we study the gravitational radiation emitted by nonradial oscillations of a neutron star excited by the mechanical impact of stochastically accreting clumps of matter. 
We calculate analytically the star's oscillatory response to a single impulsive, delta-function clump in terms of a Green's function and hence calculate the gravitational wave signal emitted by multiple delta-function or top-hat clumps, which recur periodically or with Poisson statistics. 
The mode calculations are done for a nonrotating, unmagnetised, one-component star with a polytropic EOS assuming Newtonian gravity and the Cowling approximation and reproduce standard results in the literature for the eigenfrequencies and eigenfunctions. 
The outputs of the calculations include gravitational waveforms as functions of the EOS and type and duration of impact (Figure~\ref{fig:waveforml234}), which will help to guide future work on detection algorithms, as well as predictions of the signal's amplitude and spectral properties, such as the temporal autocorrelation function [Equation~\eqref{eqn: autocorrelation_h0_scaling}], root-mean-square wave strain [Equation~\eqref{eqn: h_rms}], and ASD (Figure~\ref{fig:ASD_h0} and \ref{fig:ASD_near_LIGO_curve}), which are important for assessing detectability.

The main quantitative results are summarised as follows. 
\begin{enumerate*}[(i)]
    \item The gravitational waveforms take the form of sequences of amplitude-modulated packets, which last \(\sim 10^{-3}\)s for \(n_\text{poly}=1\) after each excitation.
    Poisson impacts lead to more irregular waveforms and slightly increase the signal strength compared to periodic impacts.
    \item The signal amplitude \(h_{\text{rms}}\) increases with \(n_{\text{poly}}\) and decreases with $n$ for \(n_{\text{poly}}=1\).
    Summing over modes with damping time \(\tau_{nl} \lesssim 10^{8}\si{yr}\) and multipole order \(2\leq l\leq4\), \(h_{\text{rms}}\) ranges from \(10^{-33}\) to \(10^{-32}\) for \(1\leq n_{\text{poly}}\leq 2\). 
    \item The $f$-mode dominates the energy spectral density \(dE^{(2,0)}/df\) in simulations by \citet{Nagar_etal2004} of transient, infalling, quadrupolar mass shells during supernova core collapse.
    In this paper, where the modes are excited continuously and stochastically by accretion, higher-$n$ modes dominate, as the spectral window lengthens.
    \item $g$-modes produce higher and narrower peaks than $f$- and $p$-modes in the ASD in the LVK observing band. The highest peaks \(\sim 10^{-26}\si{Hz}^{-1/2}\) approach the sensitivity of LVK detectors for modes with \(\tau_{nl} \sim 10^{8}\si{yr}\). 
    \item Overall, the signal is about four orders of magnitude weaker than LVK detectors can detect at present.
\end{enumerate*}

The calculations in this paper are idealised. They ignore rotation, magnetic fields, general relativistic corrections, and multiple stellar components (e.g., crust, superfluid).
To further guide searches involving real data, it is worth extending the theory to treat a realistic EOS, model the size and location of the impact region realistically, and consider toroidal oscillations such as $r$-modes \citep{Lindblom_etal1998, OwenEtAl1998}.

\section*{Acknowledgements}
This research is supported by the Australian Research Council (ARC) through the Centre of Excellence for Gravitational Wave Discovery (OzGrav) (grant number CE170100004). The authors thank Julian Carlin, Thippayawis Cheunchitra, and Kok Hong Thong for helpful discussions.

\section*{Data Availability}

Data including the numerical code used in this article are available to be shared upon request.



\bibliographystyle{mnras}
\bibliography{GWANSO} 



\appendix
\section{Oscillatory response to an impulsive impact}
\label{app:green_function_method}
In this appendix, we calculate the impulse response of the star, i.e. the mode amplitudes excited by a delta-function impact in time and space. We derive the Green's function for the problem in Appendix~\ref{app1a:green_function} and prove that surface terms vanish in Appendix~\ref{app1b:boundary_terms_in_the_general_solution}.

\subsection{Green's function}
\label{app1a:green_function}
We denote the hyperbolic operator of interest in equation~\eqref{eqn: pulsation_equation} by
\begin{align}
    \mathcal{O}_{ij} = \rho(r)\qty(\mathcal{L}_{ij} + \delta_{ij}\pdv[2]{t})
    \,,
\end{align}
where \(\delta_{ij}\) is the Kronecker delta, and \(\mathcal{L}_{ij}\) satisfies
\begin{align}
    \mathcal{L}_{ij} \delta r_j (\vb{x})
    =& - \pdv{x_{i}}(\Gamma_{1} \frac{P_0}{\rho_0} \pdv{\delta r_j}{x_{j}})
        - \pdv{x_{i}}(\frac{1}{\rho_0} \delta r_j \pdv{P_0}{x_{j}})
    \nonumber \\
    &- (\hat{\vb{r}})_{i} \left(\frac{1}{\rho_0} \dv{\rho_0}{r} - \frac{1}{\Gamma_{1} P_0}\dv{P_0}{r}\right) \Gamma_{1}\frac{P_0}{\rho_0} \pdv{\delta r_j}{x_{j}}
    \,.
    \label{eqn: L_operator}
\end{align}
In \eqref{eqn: L_operator}, the subscript $i$ denotes the $i$-th Cartesian component, and the Einstein summation convention on repeated indices is adopted.
The problem of a neutron star experiencing a spatially distributed external force density \(\mathcal{F}_k(\vb{x},t)\) can then be formulated as 
\begin{align}
    \mathcal{O}_{ki} \delta r_i(\vb{x},t) &= \mathcal{F}_k(\vb{x},t) \,.
    \label{eqn: general_component_defining}
\end{align}
The Green's function \(G_{ij}(\vb{x},t; \vb{x'},t')\) corresponding to \eqref{eqn: general_component_defining} satisfies
\begin{align}
    \mathcal{O}_{ki} G_{ij}(\vb{x},t; \vb{x'},t') &= \delta_{kj} \delta^3(\vb{x}-\vb{x'}) \delta(t-t') 
    \,,
    \label{eqn: Green_component_defining}
\end{align}
where \(G_{ij}\) is a tensor which specifies the $i$-th component of the displacement caused by the $j$-th component of the external force.
We expand the Green's function as a linear combination of the eigenfunctions \(\boldsymbol{\xi}^{(\alpha)}\) of \(\mathcal{L}_{ij}\), viz.
\begin{align}
    G_{ij}(\vb{x},t; \vb{x'},t') 
    = \sum_{\alpha,\,\beta} b_{\alpha\beta}(t,t') \xi^{*(\beta)}_{i}(\vb{x}') \xi^{(\alpha)}_{j}(\vb{x})
    \,,
    \label{eqn: G_ij_expansion}
\end{align}
where the coefficient \(b_{\alpha\beta}(t,t')\) satisfies
\begin{align}
    \qty[\sigma_{\beta}^2 + \pdv[2]{b_{\alpha\beta}(t;t')}{t}] \xi^{*(\beta)}_{k}(\vb{x}') 
    = \frac{\xi^{*(\alpha)}_{k}(\vb{x}')}{J} \delta(t-t')
    \,, \label{eqn: coefficient_b_alpha_beta}
\end{align}
and \(J=M_* R_{*}^{2}\) is a normalisation constant. 
The orthogonality condition,
\begin{align}
    \int_V d^3x \, \rho \boldsymbol{\xi}^{*(\alpha)} \vdot \boldsymbol{\xi}^{(\alpha')} 
    = J \delta_{\alpha \alpha'}
\end{align}
is applied to obtain \eqref{eqn: coefficient_b_alpha_beta}.
For \(\alpha=\beta\), \(b_{\alpha\beta}(t;t')\) is the Green's function for a one-dimensional harmonic oscillator and can be solved by Laplace transforming \eqref{eqn: coefficient_b_alpha_beta} with the initial conditions
\begin{align}
    b_{\alpha\beta}(0;t') 
    &= \pdv{b_{\alpha\beta}(t;t')}{t}\eval_{t=0} 
    \\
    &= 0 \,.
\end{align}
The result is
\begin{align}
    b_{\alpha\beta}(t, t') = \frac{\delta_{\alpha\beta} H(t-t') \sin[\sigma_{\alpha}(t-t')]
    }{J \sigma_{\alpha}}
    \label{eqn: b_alpha_beta_solution}
\end{align}
for \(\alpha=\beta\), where \(H(\cdot)\) is the Heaviside function. 
Equations~\eqref{eqn: G_ij_expansion} and \eqref{eqn: b_alpha_beta_solution} combine to yield
\begin{align}
    G_{ij}(\vb{x},t; \vb{x'},t') = H(t-t')& \sum_{\alpha} 
    \frac{\sin[\sigma_{\alpha}(t-t')]}{\sigma_{\alpha} J} 
    \xi_{i}^{*(\alpha)}(\vb{x}) \xi_{j}^{(\alpha)}(\vb{x'}) \,.
    \label{eqn: Green_component_solution}
\end{align}

\subsection{Surface terms in the general solution}
\label{app1b:boundary_terms_in_the_general_solution}
The general solution of \eqref{eqn: general_component_defining} for arbitrary \(\mathcal{F}_k (\vb{x}, t)\) is the sum of a volume integral over the stellar interior and a surface integral over the stellar boundary. The surface integral takes the form
\begin{align}
    S_{k}(\vb{x}, t) 
    & = \delta r_k(\vb{x},t) - \int d^3x' dt' \, G_{ki}^{*} \mathcal{F}_i(\vb{x}',t') 
    \label{eqn: S_k line1} 
    \\
    & = \int \, d^3x' dt' [
        \delta r_i(\vb{x'},t') \mathcal{O}_{ij} G_{jk}(\vb{x},t; \vb{x'},t')
        \nonumber \\
        & \quad - G_{ki}^{*}(\vb{x},t; \vb{x'},t') \mathcal{O}_{ij} \delta r_{j}(\vb{x'},t) ]
    \\
    & = \int d^3x' dt' \, \rho \qty(\delta r_i \mathcal{L}_{ij} G_{jk} 
                - G_{ki}^{*} \mathcal{L}_{ij} \delta r_j )
        \nonumber \\
        & \quad + \int d^3x' dt' \, \rho \qty(\delta r_i \frac{\partial^2}{\partial t^{'2}} G_{ik}
                - G_{ki}^{*} \frac{\partial^2}{\partial t^{'2}} \delta r_i) \,.
    \label{eqn: S_k line3}
\end{align}

The first line in \eqref{eqn: S_k line3} is zero, as the operator \(\mathcal{L}_{ij}\) is Hermitian under the assumption that both \(P_0\) and \(\rho_0\) vanish on the surface and both \(\grad P_0\) and \(\grad\rho_0\) are directed radially in the unperturbed state \citep{Cox1980}.

The second line in \eqref{eqn: S_k line3},
\begin{align}
    \int d^3x' \int_{0}^{t} dt' \, \rho \qty(\delta r_i \frac{\partial^2}{\partial t^{'2}} G_{ik}
            - G_{ki}^{*} \frac{\partial^2}{\partial t^{'2}} \delta r_i) \,,
    \label{eqn: S_k line3_second_line}
\end{align}
is also zero.
Integrating the temporal integral in \eqref{eqn: S_k line3_second_line} by parts,
we obtain
\begin{align} 
    \int_V d^3x' \, \rho \qty(\delta r_i \pdv{G_{ik}}{t'}\eval_{t'=0}^{t'=t} 
        - G_{ki}^{*} \pdv{\delta r_i}{t'}\eval_{t'=0}^{t'=t} ) 
    \,.
\end{align}
For \(t'=t\), equation \eqref{eqn: Green_component_solution} implies
\begin{align}
    G_{ij}(\vb{x},t; \vb{x'},t) 
    &=\frac{\partial G_{ij}(\vb{x},t; \vb{x'},t')}{\partial t'}\eval_{t'=t} 
    \\
    &= 0
    \,. \label{eqn: S_k line3_second_line_simplified}
\end{align}
Moreover, the initial values satisfy
\begin{align}
    \delta r_i(\vb{x},0) &= 0
\end{align}
and
\begin{align}
    \pdv{\delta r_i(\vb{x},t)}{t}\eval_{t=0} &= 0 \,.
\end{align}
Equation~\eqref{eqn: S_k line3_second_line_simplified} and hence equation~\eqref{eqn: S_k line1} are zero, i.e. there are no boundary terms in the formal solution. 
We then establish the equivalence
\begin{align}
    \delta r_k(\vb{x},t) 
    = \int d^3x' dt' \, G_{ki}^{*} \mathcal{F}_i(\vb{x}',t') \,.
\end{align}

\section{Numerical evaluation of the overlap integral}
\label{app:overlap_integral}

\begin{figure}
    \centering
    \includegraphics[width=\columnwidth]{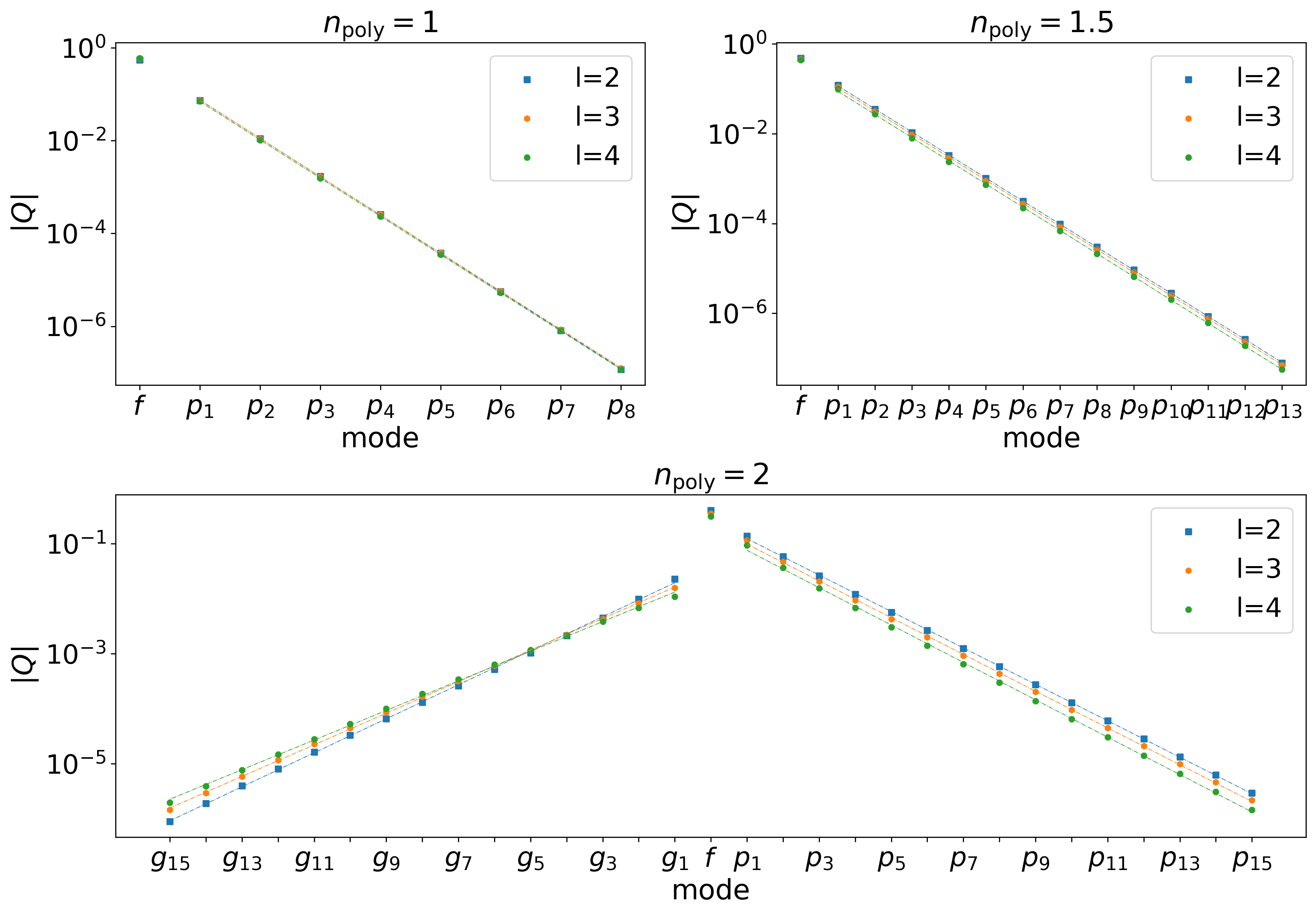}
    \caption{Overlap integral $\abs{Q_{nl}}$ for modes with \(n_{\text{poly}}=1.0\) (top left panel), 1.5 (top right panel), and 2.0 (bottom panel). The blue, orange, and green dots are for multipole orders \(l=2,3\), and 4 respectively.
    The trends in every panel are exponential (log-linear axes). Every second $p$- and $g$-mode in the bottom panel is not labelled for readability.}
    \label{fig:Q_trend}
\end{figure}

The overlap integral \(Q_{nl}\) in \eqref{eqn: Q_definition} decreases quickly as $n$ and $l$ increase. The integrand in \eqref{eqn: Q_definition} oscillates rapidly for large $n$ and $l$, making it difficult to compute \(Q_{nl}\) accurately.
Moreover, as noted by \citet{Reisenegger1994}, small mixing of $f$-modes in the numerical computation of eigenfunctions can introduce errors in \(Q_{nl}\), even if the eigenfrequencies are calculated accurately. 
If the numerical eigenfunction \(\xi^{(nlm)}\) acquires a numerical error of the form
\begin{align}
    \tilde{\xi}^{(nlm)} 
    = \qty(1 - \abs{\epsilon_0}^2 - \abs{\epsilon}^2)^{1/2} \xi^{(nlm)} 
        + \epsilon_0 \xi^{(0lm)} 
        + \epsilon \sum_{\abs{n}>0} \xi^{(nlm)} ,
\end{align}
with \(\abs{\epsilon} \ll 1\) and \(\abs{\epsilon_0} \ll 1\), the leading-order error in computing \(Q_{nl}\) of \(\tilde{\xi}^{(nlm)}\) is \(\sim \epsilon_0\), because $f$-modes mostly dominate with \(Q_{nl} \approx 0.5\) (except in the most centrally condensed stars).
Therefore, one needs \(\epsilon_0 / \abs{Q_{nl}} \ll 1\), in order to evaluate \(Q_{nl}\) accurately \citep{Reisenegger1994}. 
One way to estimate \(\epsilon_0\) is to check the orthogonality conditions \citep{Lai1994}.
In this work, we find \(\epsilon_0 \sim 10^{-9}\), which is of the same order of magnitude as the precision in solving the Lane-Emden equation.

The three panels in Figure~\ref{fig:Q_trend} are plotted on log-linear axes. We find that \(Q_{nl}\) decreases approximately exponentially with \(n\) for \(1 \leq n_{\text{poly}} \leq 2\), in line with the trend for modes with \(n\leq 4\) in \citet{PassamontiEtAl2021} for example.

\section{Intuitive spring-based model of impact}
\label{app:duration_of_collision}
In this appendix, we develop intuition about the energy and momentum transfer that occur, when a clump of matter strikes the star's surface, by analysing a toy model of an idealised, double-spring system.

\begin{figure}
    \centering
    \begin{tikzpicture}
        \draw[thick,decorate,decoration={coil,amplitude=5, segment length=5}] (0,0) -- (2,0);
        \draw[thick,decorate,decoration={coil,amplitude=5, segment length=5}] (3.5,0) -- (4.5,0);
        \filldraw[fill=blue!30] (2,-0.75) rectangle (3.5,0.75);
        \node at (2.75,0) {$M$};
        \filldraw[fill=blue!30] (4.5,-0.25) rectangle (5.0,0.25);
        \node at (4.75,0) {$m$};
        \draw[thick] (0,1.75) -- (0,-1.5);
        \foreach \y in {-1,-0.5,0,0.5,1,1.5} {
            \draw[thick] (0,\y) -- (-0.25,\y-0.25);
        }
        \node at (1,0.5) {$K$}; \node at (1,-0.5) {$R$};
        \node at (4,0.5) {$k$}; 
        \draw[<->, thick] (3.5,-0.27) -- (4.5,-0.27); \node at (4,-0.5) {$x_1$};
        \draw[<->, thick] (3.75,-1.3) -- (5.25,-1.3); \node at (4.5,-1.45) {$L$};
        \draw[<->, thick] (3.5,-0.8) -- (3.75,-0.8); \node at (3.6,-1.05) {$x_2$};
        \draw[<-, ultra thick, red] (4.45,0.9) -- (5.0,0.9);
        \node[red, font=\small] at (4.85,1.15) {$v, g$};
        \draw[dotted] (3.75,1.5) -- (3.75,-1.5);
        \draw[dotted] (5.25,1.5) -- (5.25,-1.5);
    \end{tikzpicture}
    \caption{Schematic illustration of the impact of an accreted clump on the stellar surface, modelled as an idealised mechanical system comprising two springs (spring constants $K$ and $k$) connecting two masses ($M$ and \(m \ll M\)).}
    \label{fig:spring_collision}
\end{figure}
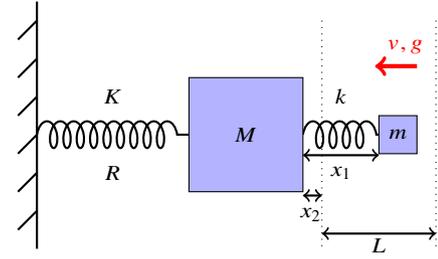

Consider a heavy spring (the neutron star) with variable length \(R\) and fixed spring constant \(K\), attached to a light spring (the clump) with initial length \(L\) and spring constant \(k\).
For simplicity, we ignore the length dependence in \(k\), which describes the deformation of the clump during an impact.
We model the system as in Figure~\ref{fig:spring_collision}, where the springs are massless but are attached to blocks of mass $M$ and \(m \ll M\), corresponding to the star and clump respectively.
The self-gravity of \(M\) is neglected.
The clump initially has a speed \(v\) and experiences an acceleration of magnitude \(g\) towards \(M\). 

Let \(x_1\) denote the variable length of the light spring and \(x_2\) denote the displacement of \(M\) from it equilibrium location. The equations of motion for this system are
\begin{align}
        \ddot{x}_1(t) &= -\frac{k}{m} [x_1(t)-x_2(t)-L] - g, 
        \label{diffeqn:x_clump}
        \\
        \ddot{x}_2(t) &= -\frac{K}{M} x_2(t) + \frac{k}{M}[x_1(t)-x_2(t)-L], 
        \label{diffeqn:x_NS}
\end{align}
with initial conditions \(x_1(0) = L, \, \dot{x}_1(0) = -v, \, x_2(0) = 0\), and \(\dot{x}_2(0) = 0\).
An overdot denotes a derivative with respect to time. 
The analytic solution of \eqref{diffeqn:x_clump} and \eqref{diffeqn:x_NS} is elementary but cumbersome. Instead, we make the approximation \(k\ll K\) and ignore the \(k/M\) term in \eqref{diffeqn:x_NS} to obtain
\begin{align}
    x_1(t) &\approx L + \omega^{-2}[-g + g \cos(\omega t) - v \omega \sin(\omega t)],
\end{align} 
with \(\omega^2 = k/m\) and \(x_2(t) \approx 0\).

Two distinct time-scales are relevant to impact duration: \(t_1>0\) such that \(x(t_1)=0\) (time to reach maximum compression), and \(t_2>0\) such that \(\dot{x}(t_2)=0\) (time to come to rest). 
Typically, for \(v \lesssim 10^{8}\si{m\,s^{-1}}\) and \(g \sim 10^{12} \si{m\,s^{-2}}\), the clump reaches minimum compression (and becomes deformed irreparably) before it comes to rest.
Therefore, it makes sense to identify \(t_1\) with the impact duration \(T\), viz.
\begin{align}
    T &= \frac{1}{\omega} 
    \arctan\qty[\frac{\omega L}{v} \frac{\frac{\omega^2v^2}{g^2} + \frac{v^2}{Lg}\qty(\sqrt{1+2gL/v^2-L^2\omega^2/v^2}-1)}{1+\frac{\omega^2v^2}{g^2}\qty(\sqrt{1+2gL/v^2-L^2\omega^2/v^2}-\frac{gL}{v^2})}] \,.
    \label{eqn: impact_duration_complicated}
\end{align}
For \(L\lesssim R_{*}\) and \(\omega \lesssim 10^4 \,\si{rad\,s^{-1}}\), we have \(L\omega/v \lesssim v\omega/g \ll 1 \), and \eqref{eqn: impact_duration_complicated} simplifies to 
\begin{align}
    T\approx \frac{L}{v} ,
\end{align}
independent of \(\omega\). 

A rough upper bound on the amplitude of the recoil of the neutron star is obtained by replacing $[x_1(t)-x_2(t)-L]$ in \eqref{diffeqn:x_NS} with \(L\). This yields
\begin{align}
    x_2(t) = \frac{m}{M}\frac{L\omega^2}{\Omega^2} \qty[\cos(\Omega t)-1] ,
\end{align}
with \(\Omega^2 = K/M\). For \(m\ll M\) and \(\omega<\Omega\), we have \(|x_2(t)| \ll L < R_*\).
Therefore, the recoil is negligible. 

\section{Shot-noise statistics for accretion impacts}
\label{app:statistics_for_the_shot_process}
In this appendix, we calculate the temporal autocorrelation function of the mode amplitude \(A_{\alpha}(t)\) excited by stochastic impacts recurring according to a shot process like \eqref{eqn: shot_process}. 
Ensemble averages for random variables proportional to \(\pdv*{A_{\alpha}(t)}{t}\) follow the same form. We perform the calculation in two equivalent ways: by modelling the impacts discretely (Appendix~\ref{appD1:discrete_impacts}) and continuously (Appendix~\ref{appD2:Equivalent approach in the continuous limit}).
Both approaches yield the same results.

\subsection{Discrete impacts}
\label{appD1:discrete_impacts}
The total mode amplitude \(A_{\alpha}(t)\) is related to \(A_{\alpha}(t; t_{\text{s}})\) defined by \eqref{eqn: A_alpha(t;t_s)} via
\begin{align}
    A_{\alpha}(t) = \sum_{\{t_\text{s}\}} A_{\alpha}(t; t_{\text{s}}) ,
    \label{eqn: A_discrete}
\end{align}
where \(\{t_\text{s}\}\) is a set of impact times that are Poisson distributed with time-independent mean rate \(\lambda=f_{\text{acc}}\).
We assume that \(A_{\alpha}(t; t_{\text{s}})\) is a stationary random variable.
The expected value of \(A_{\alpha}(t)\) is defined by
\begin{align}
    &\expval{A_{\alpha}(t)} 
    = \int d\{t_\text{s}\} \, p(\{t_\text{s}\}) A_{\alpha}(t) \,,
    \label{eqn: <A>_discrete_defn}
\end{align}
where \(p(\cdot)\) denotes the probability density function, and the dimension of the integral in \eqref{eqn: <A>_discrete_defn} equals the cardinality of \(\{t_{\text{s}}\}\).
In order to compute \eqref{eqn: <A>_discrete_defn}, we perform a weighted sum over $n$ impacts before time $t$ and set the lower bound of \(t_{\text{s}}\) to be zero without loss of generality. The result is
\begin{align}
    \expval{A_{\alpha}(t)} 
    =& \sum_{n=1}^{\infty}  
        \int_0^{t_{\text{s}2}} dt_{\text{s}1} 
        \int_0^{t_{\text{s}3}} dt_{\text{s}2}... \int_0^{t} dt_{\text{s}n} \,
        p(t_{\text{s}1},...,t_{\text{s}n})
        \nonumber \\ &\times
        \text{Pr}[N(t-t_{\text{s}n})=0]
        \sum_{i=1}^{n} A_{\alpha}(t; t_{\text{s}i})
        \label{eqn: D3}
    \\
    =& \sum_{n=1}^{\infty} 
        \frac{1}{n!} \int_0^{t} \,dt_{\text{s}1}
        \int_0^{t} dt_{\text{s}2}... \int_0^{t} \,dt_{\text{s}n} \,
        \nonumber \\ &\times
        p(t_{\text{s}1},...,t_{\text{s}n}) \, 
        \text{Pr}[N(t-t_{\text{s}n})=0]
        \sum_{i=1}^{n} A_{\alpha}(t; t_{\text{s}i})
        \label{eqn: D4}
    \\
    =& \sum_{n=1}^{\infty} 
        \frac{\lambda^n \exp(-\lambda t)}{n!} 
        \times n t^{n-1} \int_0^{t} dt_s \expval{A_{\alpha}(t; t_{\text{s}})}
        \label{eqn: D5}
    \\
    =& \, \lambda \int_0^{t} dt_s \expval{A_{\alpha}(t; t_{\text{s}})} \,,
    \label{eqn: <A>_discrete_result}
\end{align}
where Pr($\cdot$) is the probability mass function, and \(N(\Delta t)\) denotes the number of events during a length of time \(\Delta t\). 
In \eqref{eqn: D3}, \(p(t_{\text{s}1},...,t_{\text{s}n})\text{Pr}[N(t-t_{\text{s}n})=0]\) equals the probability of \(n\) events occurring in the specific order \(0 \leq t_{\text{s}1} < t_{\text{s}2} < \dots < t_{\text{s}n} \leq t\).
From \eqref{eqn: D3} to \eqref{eqn: D4}, we convert the time-ordered sequence of integrals to $n$ integrals on the interval \([0,t]\), which are insensitive to time-ordering. 
Equation~\eqref{eqn: D5} easily follows from \eqref{eqn: D4} after substituting \(p(t_{\text{s}1},...,t_{\text{s}n})\text{Pr}[N(t-t_{\text{s}n})=0] = \lambda^n \exp(-\lambda t)\).

The autocorrelation function \(\expval{A_{\alpha}(t) A_{\alpha'}(t')}\) with respect to the same set \(\{t_\text{s}\}\), assuming \(t<t'\) without loss of generality, can be expressed as
\begin{align}
    &\expval{A_{\alpha}(t)A_{\alpha'}(t')} 
    \nonumber \\
    =& \int d\{t_\text{s}\} \, p(\{t_\text{s}\}) 
        \expval{\sum_{s\in \{t_\text{s}\}} A_{\alpha}(t; s) 
                \sum_{s'\in \{t_\text{s}\}} A_{\alpha}(t'; s')}
    \\
    =& \int d\{t_\text{s}\} \, p(\{t_\text{s}\}) 
        \expval{\sum_{s\in \{t_\text{s} | t_\text{s}\leq t\}} A_{\alpha}(t; s) 
                \sum_{s'\in \{t_\text{s} | t_\text{s}\leq t\}} A_{\alpha}(t'; s')}
        \nonumber \\ &+
        \int d\{t_\text{s}\} \, p(\{t_\text{s}\})  
        \expval{\sum_{s\in \{t_\text{s} | t_\text{s}\leq t\}} A_{\alpha}(t; s) 
                \sum_{s'\in \{t_\text{s} | t< t_\text{s}\leq t'\}} A_{\alpha}(t'; s')}
    \label{eqn: autocorrelation_A_discrete_stage1}
\end{align}
Since the interval \([0, t]\) does not overlap with the interval \((t,t']\), the two sums in the second line of \eqref{eqn: autocorrelation_A_discrete_stage1} are independent statistically. 
Upon decomposing the double sums in the first line of \eqref{eqn: autocorrelation_A_discrete_stage1} into \(\sum_{s=s'} + \sum_{s\neq s'}\), and repeating the logic from \eqref{eqn: <A>_discrete_defn} to \eqref{eqn: <A>_discrete_result}, we obtain
\begin{align}
    \expval{A_{\alpha}(t)A_{\alpha'}(t')} 
    =& \, \lambda^2 \int_{0}^{t} dt_{\text{s}} \int_{0}^{t'} dt'_{\text{s}}
            \expval{A_{\alpha}(t; t_{\text{s}})} 
            \expval{A_{\alpha'}(t'; t'_{\text{s}})}
        \nonumber \\ &+
        \lambda \int_{0}^{t} dt_{\text{s}} \, \expval{A_{\alpha}(t; t_{\text{s}}) A_{\alpha'}(t'; t_{\text{s}})}
    \label{eqn: <AA>_discrete} \,,
\end{align}
with
\begin{align}
    &\expval{A_{\alpha}(t; t_{\text{s}}) A_{\alpha'}(t'; t_{\text{s}})}
    \nonumber \\ =& \,
    \abs{ \frac{\vb{p} \vdot \boldsymbol{\xi}^{*(\alpha)}(\vb{x}_\text{f})}{J} }^2
    \int ds \int ds' \, g_{\alpha}(t-s) g_{\alpha'}(t'-s') 
    \nonumber \\ &\times
    \expval{ \mathcal{T}(s-t_{\text{s}}) 
    \mathcal{T}(s'-t_{\text{s}}) }
    \,.
    \label{eqn: D10}
\end{align}
Equation~\eqref{eqn: D10} follows from \eqref{eqn: A_alpha(t;t_s)} in the main text, and \(\mathcal{T}(t)\) is defined by \eqref{eqn: tophat}.

\subsection{Equivalent approach in the continuous limit}
\label{appD2:Equivalent approach in the continuous limit}
When the impact rate is high, the discrete events at \(\{t_{\text{s}} \}\) can be approximated by a continuum of events, an infinitesimal number of which occur per infinitesimal time interval. We can recalculate \(\expval{A_{\alpha}(t) A_{\alpha'}(t')}\) in the continuous limit to check the results in Appendix~\ref{appD1:discrete_impacts}, including \eqref{eqn: <AA>_discrete}.
The following approach is adapted from \citet{cox1977theory}. 

The continuous analog of \eqref{eqn: A_discrete} is
\begin{align}
    A_{\alpha}(t) = 
    \int_{-\infty}^t \int_{Q} \, dN(t_{\text{s}}, Q) 
    A_{\alpha}(t; t_{\text{s}}, Q) \,,
    \label{eqn: A_cts}
\end{align}
where \(dN(t,Q)\) equals the number of collisions in the interval $(t,t+dt)$ associated with another continuous random variable \(Q\) (to keep the calculation general), whose value lies within \((Q, Q+dQ)\).

We restrict attention to a point process without simultaneous events. We also assume that the probability density function of \(Q\) at time \(t\) is \(q(Q;t) = q(Q)\), i.e. \(Q\) is a time-independent random variable. 
The Poisson shot process obeys
\begin{align}
    \expval{dN(t,Q)} 
    &= \text{var}[dN(t,Q)] \\
    &= \lambda q(Q) dt dQ ,
\end{align}
which implies 
\begin{align}
    \expval{A_{\alpha}(t)}
    =& \int_{-\infty}^t dt_{\text{s}} \int dQ 
        \lambda q(Q) A_{\alpha}(t; t_{\text{s}}, Q) 
    \\
    =& \, \lambda \int_{-\infty}^t dt_{\text{s}} \, \expval{A_{\alpha}(t; t_{\text{s}})}_{Q} ,
    \label{eqn: <A>}
\end{align}
where \(\expval{\cdot}_{Q}\) denotes an ensemble average over \(Q\).

The autocorrelation function can be derived by examining
\begin{align}
    &\expval{dN(t,Q_1) dN(s,Q_2)} 
    \nonumber \\
    =& \, \text{Pr}[dN(t)=1, dN(s)=1] q(Q_1, Q_2; t, s) dQ_1 dQ_2 \\
    =& \, [\lambda w(s-t) q(Q_1) q(Q_2)
        \nonumber \\ &+ 
        \lambda q(Q_1) \delta(s-t)\delta(Q_1-Q_2) ] dt ds dQ_1 dQ_2 
    \,. \label{eqn: <dN dN>}
\end{align}
In equation~\eqref{eqn: <dN dN>},
\begin{align}
    w(s-t) = \lim_{\Delta s\rightarrow 0} \frac{\text{Pr[event in $(s,s+\Delta s)|$ event at \(t\)]}}{\Delta s}
\end{align}
is a function generalising the renewal density in \citet{cox1977theory} [see Sec. 9, Equations~(1) and (64) in the latter reference]. For a Poisson process, we have \(w(s-t) = \lambda\). 
Upon combining \eqref{eqn: A_cts} and \eqref{eqn: <dN dN>}, we arrive at
\begin{align}
    \expval{A_{\alpha}(t) A_{\alpha'}(t')} 
    =& \, \lambda^2 \int_{-\infty}^t dt_{s} \int_{-\infty}^{t'} dt'_{s} \,
            \expval{A_{\alpha}(t; t_{\text{s}})}_{Q} 
            \expval{A_{\alpha'}(t'; t'_{\text{s}})}_{Q}
        \nonumber\\ &+
        \lambda \int_{-\infty}^t dt_{s} \,
            \expval{A_{\alpha}(t; t_{\text{s}}) A_{\alpha'}(t'; t_{\text{s}})}_{Q}
    ,
    \label{eqn: <AA>_cts}
\end{align}
which is equivalent to \eqref{eqn: <AA>_discrete}.


\bsp	
\label{lastpage}
\end{document}